\documentclass[a4paper,onecolumn,10pt,accepted=2025-11-25]{quantumarticle}
\pdfoutput=1
\usepackage{geometry}
 \usepackage[utf8]{inputenc}
\usepackage[english]{babel}
\usepackage[T1]{fontenc}
\usepackage{graphicx,graphics,times,bm,bbm,bbold,amssymb,soul}
\usepackage{amsthm,mathtools,amsmath,mathrsfs,amsfonts,dsfont}
\usepackage{color,xcolor,dcolumn,hyperref,cleveref,lipsum}
\usepackage{fix-cm,nicematrix,tikz,relsize,yfonts,enumitem}
     
\newcolumntype{P}[1]{>{\centering\arraybackslash}p{#1}}

\usepackage{tabularx,ragged2e}

\DeclareMathOperator{\bk}{\mathbf{k}}
\DeclareMathOperator{\bs}{\mathbf{s}}
\DeclareMathOperator{\Span}{Span}
\DeclareMathOperator{\Sym}{Sym}

\newtheorem{theorem}{Theorem}

\newtheorem{manuallemmainner}{Lemma}
\newenvironment{manuallemma}[1]{%
  \renewcommand\themanuallemmainner{#1}%
  \manuallemmainner
}{\endmanuallemmainner}

\newtheorem{manualremarkinner}{Remark}
\newenvironment{manualremark}[1]{%
  \renewcommand\themanualremarkinner{#1}%
  \manualremarkinner
}{\endmanualremarkinner}

\newtheorem{manualpropositioninner}{Proposition}
\newenvironment{manualproposition}[1]{%
  \renewcommand\themanualpropositioninner{#1}%
  \manualpropositioninner
}{\endmanualpropositioninner}

\newtheorem{manualdefinitioninner}{Definition}
\newenvironment{manualdefinition}[1]{%
  \renewcommand\themanualdefinitioninner{#1}%
  \manualdefinitioninner
}{\endmanualdefinitioninner}

\newtheorem{manualclaiminner}{Claim}
\newenvironment{manualclaim}[1]{%
  \renewcommand\themanualclaiminner{#1}%
  \manualclaiminner
}{\endmanualclaiminner}

\newtheorem{manualexampleinner}{Example}
\newenvironment{manualexample}[1]{%
  \renewcommand\themanualexampleinner{#1}%
  \manualexampleinner
}{\endmanualexampleinner}

\usepackage{ulem,cancel}

\begin{document}
%%%%%%%%%%%%%%%%%%%%%%%%%%%%%%%%%%%%%%%%%%%%%%%%%
%%%%%%%%%%%%%%%%%%%%%%%%%%%%%%%%%%%%%%%%%%%%%%%%%

\title{Entangled Subspaces through Algebraic Geometry}

\author{Masoud Gharahi}
\email[]{masoud.gharahi@gmail.com}
\email[]{masoud.gharahighahi@units.it}
\affiliation{Faculty of Physics, Astronomy and Applied Computer Science, Jagiellonian University, 30-348 Kraków, Poland}
\affiliation{Department of Physics, University of Trieste, Strada Costiera 11, 34151 Trieste, Italy}
\orcid{0000-0002-7515-6179}
\homepage{https://sites.google.com/view/masoudgharahi}

\author{Stefano Mancini}
\email[]{stefano.mancini@unicam.it}
\affiliation{School of Science and Technology, University of Camerino, 62032 Camerino, Italy}
\affiliation{INFN Sezione di Perugia, 06123 Perugia, Italy}

%%%%%%%%%%%%%%%%%%%%%%%%%%%%%%%%%%%%%%%%%%%%%%%%%

\begin{abstract}
We propose an algebraic geometry-inspired approach for constructing entangled subspaces within the Hilbert space of a multipartite quantum system. Specifically, our method employs a modified Veronese embedding, restricted to the conic, to define subspaces within the symmetric part of the Hilbert space. By utilizing this technique, we construct the minimal-dimensional, non-orthogonal yet Unextendible Product Basis (nUPB), enabling the decomposition of the multipartite Hilbert space into a two-dimensional subspace, complemented by a Genuinely Entangled Subspace (GES) and a maximal-dimensional Completely Entangled Subspace (CES). In multiqudit systems, we determine the maximum achievable dimension of a symmetric GES and demonstrate its realization through this construction. Furthermore, we systematically investigate the transition from the conventional Veronese embedding to the modified one by imposing various constraints on the affine coordinates, which, in turn, increases the CES dimension while reducing that of the GES.
\end{abstract}

\maketitle

%%%%%%%%%%%%%%%%%%%%%%%%%%%%%%%%%%%%%%%%%%%%%%%%%

\section{Introduction}
Once entanglement was recognized as a valuable resource for quantum information processing, identifying entangled states in multipartite systems became a crucial objective \cite{HHHH09}. Although entanglement is traditionally understood as a property of individual quantum states, it can also be extended to entire subspaces of the Hilbert space. This broader viewpoint facilitates a more holistic study of entanglement, focusing not just on states, but on entire subspaces where all states are entangled. A central notion in this context is the Completely Entangled Subspace (CES), defined as a subspace where no vector is fully separable \cite{BDMSST99, Parthasarathy04, Bhat, WS08, ATL11, Johnston13, SAS14, BC18}. The study of CESs thus generalizes the concept of entanglement, providing a more comprehensive framework for understanding quantum correlations within the Hilbert space. One of the most well-established and structured methods for constructing CESs relies on Unextendible Product Bases (UPBs) \cite{BDMSST99, AL01, DMSST03, Pittenger, Johnston14}. A UPB is a finite set of product states that spans a proper subspace of the Hilbert space, with the unique property that no additional product state exists orthogonal to all members of the set. The absence of extra orthogonal product states implies that the orthogonal complement (orthocomplement) of the subspace spanned by the UPB contains only entangled states, thus forming a CES. This connection provides an effective and operational framework for generating CESs and has significant implications for quantum information theory. In particular, the interplay between UPBs and CESs has greatly enhanced our understanding of the fundamental properties of entanglement \cite{BDMSST99, Parthasarathy04, Bhat, WS08, ATL11, Johnston13, SAS14, BC18, AL01, DMSST03, Pittenger, Johnston14}.

Although fully separable states are, by definition, excluded from a CES, states with partial separability may still reside within it. This consideration motivates the notion of subspaces that exclude all forms of partial separability, thereby containing only Genuinely Multipartite Entangled (GME) states (also known as indecomposable states). In Ref.~\cite{DA18}, the authors referred to such spaces as Genuinely Entangled Subspaces (GESs). Since a GES contains only GME states, the construction of such subspaces is of significant interest. In particular, identifying explicit bases for GESs is especially valuable, as multipartite entanglement manifests in various inequivalent forms \cite{DVC00}, making its structure highly nontrivial. However, only a limited number of construction methods are currently known. In Ref.~\cite{WS08}, it was shown that random subspaces of specific dimensions within the Hilbert space are almost surely devoid of fully separable states. More interestingly, in Refs.~\cite{HLW04, Hayden04}, it was demonstrated that, with high probability, a random subspace of dimension close to that of the Hilbert space typically consists almost entirely of generic states, i.e., nearly-maximally entangled states. While such random constructions highlight the prevalence of entanglement, explicit constructions offer a deeper understanding by enabling detailed analysis of the entanglement properties of states supported within these subspaces. In Refs.~\cite{DA18, DA20, Demianowicz22}, the authors use the concept of non-orthogonal Unextendible Product Bases (nUPBs) to construct GESs as the orthocomplement of nUPBs. The definition of an nUPB parallels that of a UPB, with the crucial distinction that the orthogonality condition among the product states is dropped \cite{Parthasarathy04,Bhat,Pittenger,LMS10,Skowronek11}. An alternative approach was proposed in Ref.~\cite{AHB19}, where the notion of unextendible biseparable bases was introduced to construct GESs that are bidistillable—meaning every state supported on the subspace exhibits distillable entanglement across all bipartitions. In addition to these approaches, Refs.~\cite{WD19, MA20, MKA23} employ the stabilizer formalism to assess whether a given stabilizer subspace qualifies as a GES. Further contributions include several methods for constructing CESs and GESs via quantum channels \cite{Antipin21}, as well as a technique that involves combining bipartite CESs using tensor products followed by joining adjacent subsystems \cite{Antipin22}. A related algebraic-geometric perspective on entangled subspaces, focusing on subspaces avoiding states of bounded border rank, was developed in Ref.~\cite{LJ22}.

The concepts of CES and GES not only provide deep insights into the structural complexity of multipartite entanglement, but have also been proven to be valuable tools in various quantum information processing tasks, such as quantum error correction \cite{GW07, HG20} and quantum cryptography \cite{SS19}. In light of their theoretical and practical significance, the development of systematic methods for constructing CESs and GESs, particularly those with explicit bases that elucidate the types of entanglement they support, remains a challenging and actively pursued objective within the field of quantum information science.

Among the various approaches proposed for CESs and GESs, UPBs and their generalization, nUPBs, stand out as systematic methods. While nUPBs have predominantly been employed in the construction of GESs, we exploit them here to construct CESs as their orthocomplement subspaces. This shift in perspective is partly motivated by the fact that nUPBs do not require pairwise orthogonality, making them more flexible and often easier to handle. Additionally, it is noteworthy that nontrivial UPBs do not exist in the Hilbert spaces 
$\mathbbm{C}^2\otimes\mathbbm{C}^d$ for $d\geq2$ \cite{BDMSST99}, and UPBs of certain sizes are also impossible \cite{Johnston14}.

The primary objective of this work is to minimize the size of an nUPB, thereby maximizing the dimension of the associated CES. To accomplish this, we propose an approach based on algebraic geometry. By encoding each product state with a single homogeneous coordinate, basis elements are represented as monomials in a single variable rather than multivariable ones. This reformulation highlights the role of highly symmetric subspaces and motivates the use of the Veronese embedding \cite{Harris} as a suitable mathematical framework. In fact, the concept of the nUPB is inherently tied to algebraic geometry when viewed through the lens of the Veronese embedding. This framework offers valuable insights into the structure and characterization of nUPBs.

However, the standard Veronese embedding is not sufficient to achieve the CES of maximal dimension in systems beyond multiqubit. To address this, we generalize the approach to arbitrary $n$-partite systems by considering the image of the Veronese multidegree embedding applied to $n$ copies of a single qubit, followed by a projective map to eliminate redundant coordinates. This method allows us to decompose the $n$-partite Hilbert space into a two-dimensional subspace, complemented by a GES and a CES of maximal possible dimension. Our modified Veronese embedding imposes the maximal set of constraints on the coordinates of the nUPB points, resulting in a refined Segre–Veronese embedding that effectively minimizes its dimensionality. Additionally, by applying an orthogonalization process, such as the Gram-Schmidt procedure, we can transform the nUPB set into a collection of pairwise orthogonal states. Due to the symmetry of the nUPB, we observe that after orthogonalization, a significant subset of the nUPB can be transformed into a GES. For multiqudit systems, we determine the maximum size of a symmetric GES and establish that it can be achieved through our construction. Furthermore, we systematically examine the transition from the standard Veronese embedding to its modified form by imposing various constraints on the affine coordinates. This procedure increases the dimension of the CES while correspondingly reducing that of the GES. Additionally, we show that multiple mutually orthogonal GESs can be extracted from the CES. These results demonstrate the effectiveness of our construction and its broader relevance for understanding multipartite entanglement through algebraic and geometric techniques.

The structure of the paper is as follows. In Section \ref{Sec.ii}, we present the necessary preliminary concepts. Section \ref{Sec.iii} introduces our algebraic-geometric framework for analyzing the multiqubit case. In Section \ref{Sec.iv}, we extend this analysis to the multiqudit setting, starting with the Veronese embedding and demonstrating that our construction allows us to attain the maximal-dimensional symmetric GES. We then propose a modified version of Veronese embedding, actually a modified version of Segre-Veronese embedding, and explore the transition between these two embeddings. This transition leads to an increase in the dimension of the CES while reducing the dimension of the GES. In Section \ref{sec.v}, we further generalize our approach to multipartite systems with heterogeneous local dimensions. Finally, Section \ref{Sec.vi} concludes the paper with a summary of our findings and a discussion of future directions.

%%%%%%%%%%%%%%%%%%%%%%%%%%%%%%%%%%%%%%%%%%%%%%%%

\section{Preliminaries}\label{Sec.ii}

In what follows, we shall be concerned with $n$-partite pure quantum states
\begin{equation}\label{n-partite}
|\psi\rangle=\sum_{j=1}^{n}\sum_{i_j\in\mathbbm{Z}_{d_j}}\mathbf{c}_{i_1\cdots{i_n}}|i_1\cdots i_n\rangle\,,
\end{equation}
which are order-$n$ tensors in finite dimensional product Hilbert spaces $\mathcal{H}_{\delta}=\otimes_{j=1}^n\mathcal{H}_{d_j}=\otimes_{j=1}^n\mathbbm{C}^{d_j}$, with $\delta=d_1\times\cdots\times d_n$ and each $d_j$ standing for the dimension of the local Hilbert space associated with the $j$-th subsystem. In this context, the set $\{|i_1\cdots i_n\rangle\equiv|i_1\rangle\otimes\cdots\otimes|i_n\rangle\mid{i_j}\in\mathbbm{Z}_{d_j},~1\leq j\leq n\}$ constitutes the canonical (also called standard or computational) basis of $\mathcal{H}_{\delta}$. In the special case where $d_j=d$ for all $1\leq j\leq n$, we denote the Hilbert space more compactly as $\mathcal{H}_{d^n}=\otimes^n\mathbbm{C}^d$.

\begin{manualdefinition}{1}\label{Def1}
An $n$-partite pure state in Eq. \eqref{n-partite} is said to be fully product (or fully separable) if it can be expressed as a simple tensor. That is, there exist pure states $|\varphi_j\rangle\in\mathcal{H}_{d_j}$ for all $1\leq j\leq n$ such that
\begin{equation}
|\psi\rangle=|\varphi_1\rangle\otimes\cdots\otimes|\varphi_n\rangle\,.
\end{equation}
Otherwise, the state $|\psi\rangle$ is called entangled.
\end{manualdefinition}

\begin{manualdefinition}{2}\label{Def2}
An $n$-partite entangled state that cannot be factorized into a biproduct (biseparable) state for any nontrivial bipartition, i.e.,
\begin{equation}
|\psi\rangle\neq|\phi\rangle_{S}\otimes|\chi\rangle_{\Bar{S}}\,, \qquad \forall~S,\Bar{S} \quad \text{s.t.} \quad S,\Bar{S}\neq\emptyset\,, \quad \text{and} \quad S\cup\Bar{S}=\{1,\ldots,n\}\,,
\end{equation}
is referred to as a Genuinely Multipartite Entangled (GME) state, or equivalently, an indecomposable state.
\end{manualdefinition}

To determine whether a multipartite state is GME, we can employ techniques from multilinear algebra. In multilinear algebra, the vectorization of an order-$n$ tensor, as in Eq.\eqref{n-partite}, is a form of tensor reshaping. In this paper, we consider a specific type of tensor reshaping—tensor flattening (or matricization) \cite{Landsberg}—which involves partitioning the $n$-fold tensor product space $\mathcal{H}_{\delta}$ into twofold tensor product spaces with enlarged local dimensions. To formalize this partitioning, we define an ordered $\ell$-tuple $I=(j_1,\ldots,j_{\ell})$, where $1\leq\ell\leq\lfloor\frac{n}{2}\rfloor$ and $1\leq j_1<\cdots<j_{\ell}\leq{n}$. The complementary partition is represented by the ordered $(n-\ell)$-tuple $\bar{I}$, such that $I\cup\bar{I}=\{1,2,\ldots,n\}$. Consequently, the Hilbert space will be written as $\mathcal{H}_{\delta}\simeq\mathcal{H}_{I}\otimes\mathcal{H}_{\bar{I}}$, where $\mathcal{H}_{I}=\otimes_{\iota=j_1}^{j_{\ell}}\mathbbm{C}^{d_{\iota}}$ and $\mathcal{H}_{\bar{I}}$ is the complementary Hilbert space. Using Dirac notation, the matricization of $|\psi\rangle\in\mathcal{H}_{\delta}$ reads $\mathcal{M}_{I}[\psi]=\left(\langle{e_1}|\psi\rangle,\ldots,\langle{e_{d_{I}}}|\psi\rangle\right)^{\rm{T}}$, where $\{|e_m\rangle=|i_{j_1}\cdots{i_{j_\ell}}\rangle\mid{i_{j_k}}\in\mathbbm{Z}_{d_{j_k}},~1\leq k\leq\ell\}_{m=1}^{d_I}$ is the canonical basis of $d_I$-dimensional Hilbert space $\mathcal{H}_{I}$, with $d_I=\prod_{\iota=j_1}^{j_{\ell}}d_{\iota}$, and ${\rm{T}}$ denotes the matrix transposition. Thus, for a given state $|\psi\rangle$, there exist as many matrix representations $\mathcal{M}_{I}[\psi]$ as the number of possible $\ell$-tuples $I$, which is ${\binom{n}{\ell}}$. This allows us to define the $\ell$-multilinear rank ($\ell$-multirank) of $|\psi\rangle$ as the ${\binom{n}{\ell}}$-tuple of ranks of the $\mathcal{M}_{I}[\psi]$ \cite{GMO20,GM21}. A Mathematica codebase for computing $\ell$-multiranks of multiqudit states is provided in Ref.~\cite{Gharahi25}.

\begin{manualremark}{1}\label{remark1}
A state is GME iff all $\ell$-multiranks are strictly greater than one.
\end{manualremark}

\begin{manualexample}{1}\label{exmp1}
For instance, the well-known GHZ and W states \cite{GHZ89,DVC00} for four qubits, defined by $|{\rm{GHZ}}_4\rangle=|0000\rangle+|1111\rangle$ and $|{\rm{W}}_4\rangle=|0001\rangle+|0010\rangle+|0100\rangle+|1000\rangle$, are both GME since their $1$- and $2$-multiranks are $(2222)$ and $(222)$, respectively. In contrast, the four-qubit state $|{\rm{BB}}\rangle=|0000\rangle+|0011\rangle+|1100\rangle+|1111\rangle$ is not GME. Although it has the same 1-multirank $(2222)$, its $2$-multirank is $(144)$, indicating separability between the first two and last two qubits. Indeed, it can be written as a biseparable state $|{\rm{BB}}\rangle=(|00\rangle+|11\rangle)_{12}\otimes(|00\rangle+|11\rangle)_{34}$ \cite{GMO20}.
\end{manualexample}

In what follows, we present the formal definitions of completely entangled subspaces and (non-orthogonal) unextendible product bases, which serve as fundamental constructs underpinning the theoretical framework developed in this work.

\begin{manualdefinition}{3}
%{(Ref.~\cite{Parthasarathy04})}
\label{Def3}
A subspace $\mathcal{C}\subsetneq\mathcal{H}_{\delta}$ is called a Completely Entangled Subspace (CES) iff it contains no fully separable state.
\end{manualdefinition}

\begin{manualdefinition}{4}
%{(Ref.~\cite{BDMSST99})}
\label{De4}
A set of pairwise orthogonal fully product vectors $\{|\psi_l\rangle\equiv\otimes_{j=1}^{n}|\varphi_l^j\rangle\}_{l=1}^{u}$ that spans a proper subspace of $\mathcal{H}_{\delta}$, i.e., $u<\dim\mathcal{H}_{\delta}$, is called Unextendible Product Basis (UPB) if its orthocomplement subspace is a CES.
\end{manualdefinition}

An analogous definition can be obtained by dropping the orthogonality condition.

\begin{manualdefinition}{5}
%{(Ref.~\cite{BDMSST99})}
\label{Def5}
A set of fully product vectors $\{|\psi_l\rangle\equiv\otimes_{j=1}^{n}|\varphi_l^j\rangle\}_{l=1}^{u}$, not necessarily pairwise orthogonal, that spans a proper subspace of $\mathcal{H}_{\delta}$ is called non-orthogonal Unextendible Product Basis (nUPB) if its orthocomplement subspace is a CES.
\end{manualdefinition}

It is worth noting that while the orthocomplement of a subspace spanned by a UPB or an nUPB is a CES, the orthocomplement of a CES is not necessarily a subspace spanned by a UPB or an nUPB \cite{WS08,Skowronek11}. Thus, the presence of a CES does not necessarily imply that its orthocomplement can be spanned by any form of UPB (or nUPB). Of course, the dimension of the subspace containing entangled states plays a crucial role. The maximum size of a CES is given by the following theorem; we recall it here for completeness since it is a crucial step for the algebraic-geometric extension developed below.

\begin{theorem}
%{(Refs.~\cite{Wallach02,Parthasarathy04})}
\label{theorem-CES}
In a Hilbert space $\mathcal{H}_{\delta}=\otimes_{j=1}^n\mathbbm{C}^{d_j}$, the maximum dimension of a CES is given by
\begin{equation}\label{max-CES}
\prod_{j=1}^{n}d_j-\sum_{j=1}^{n}(d_j-1)-1\,.
\end{equation}
\end{theorem}
{\it Proof.}
A detailed proof can be found in Refs.~\cite{Parthasarathy04,Wallach02}.
\qed

While a CES inherently excludes fully separable states, it may still contain partially separable states. This observation leads to the concept of subspaces that exclude all types of partial separability. The formal definition of such a subspace is presented below.

\begin{manualdefinition}{6}
%{(Refs.~\cite{Parthasarathy04,DA18})}
\label{Def6}
A subspace $\mathcal{G}\subsetneq\mathcal{H}_{\delta}$ is called a Genuinely Entangled Subspace (GES) iff it contains only GME states.
\end{manualdefinition}

It is worth noting that, by definition, every GES is a special case of a CES. This highlights the fact that while all GESs share the property of containing only GME states, they represent a more restrictive subset within the broader class of CESs. Consequently, understanding the properties and constraints of GESs is essential for a more detailed exploration of entanglement in multipartite systems.

The dimension of a GES is fundamental to understanding its structure. The upcoming theorem delivers a rigorous characterization of the maximum possible dimension of a GES, providing important insights into the structural limits of GESs in the context of quantum entanglement. We note that this statement was previously observed in Ref.~\cite{DA18}, which draws on results from Ref.~\cite{CMW08}. In the present work, we provide a complete and rigorous proof.

\begin{theorem}\label{theorem-GES}
In a Hilbert space $\mathcal{H}_{\delta}=\otimes_{j=1}^n\mathbbm{C}^{d_j}$, with $2\leq d_j\leq d_{j+1}$, the maximum dimension of a GES is given by
\begin{equation}\label{max-GES}
(d_1-1)(\prod_{j=2}^{n}d_j-1)\,.
%\prod_{j=1}^{n}d_j-\prod_{j=2}^{n}d_j-d_1+1\,.
\end{equation}
\end{theorem}
{\it Proof.}
We invoke Remark \ref{remark1} alongside Theorem 11 from Ref.~\cite{CMW08}, which states that the maximum dimension of a subspace with Schmidt rank at least $r$ in $\mathbbm{C}^{d_I}\otimes\mathbbm{C}^{d_{\bar{I}}}$ is given by $(d_I-r+1)(d_{\bar{I}}-r+1)$. By considering all possible partitions of the $n$-fold tensor product space $\mathcal{H}_{\delta}$ into $2$-fold tensor product spaces $\mathcal{H}_{I}\otimes\mathcal{H}_{\bar{I}}$, we can determine the maximum dimension of subspaces with Schmidt rank at least $r$ in each $\mathcal{H}_{I}\otimes\mathcal{H}_{\bar{I}}$. Furthermore, for any partition $I$, the rank of the corresponding flattening is identical to the rank of the reduced density matrix obtained after tracing over the parties
identified by the complementary partition $\bar{I}$, which is equal to the Schmidt rank \cite{GMO20,GM21}. Setting $r=2$, we ensure that any state in these subspaces is not biseparable with respect to the given bipartition $\mathcal{H}_{I}\otimes\mathcal{H}_{\bar{I}}$. To construct a GES, one must consider all subspaces with Schmidt rank at least $2$ across all possible bipartitions $\mathcal{H}_{I}\otimes\mathcal{H}_{\bar{I}}$. Notably, the projections of any two subspaces with Schmidt rank at least $2$, defined with respect to different bipartitions, have an empty intersection in the ambient projective space \cite[Theorem 1.22]{Shafarevich}. This guarantees that the dimension of a GES is not counted multiple times. Consequently, the maximum achievable dimension of a GES is the minimum of the expression $(d_I-1)(d_{\bar{I}}-1)$, taken over all bipartitions of the Hilbert space $H_{\delta}$. Finally, taking into account the constraint $2\leq d_j\leq d_{j+1}$, the result in Eq. \eqref{max-GES} follows.
\qed

%%%%%%%%%%%%%%%%%%%%%%%%%%%%%%%%%%%%%%%%%%%%%%%%%

\section{Multiqubit case}\label{Sec.iii}

As a starting point, the procedure is outlined in the simplest case of an $n$-qubit system, whose associated Hilbert space is given by \mbox{$\mathcal{H}_{2^n}=\otimes^n\mathbbm{C}^2$}. We proceed by formulating a statement whose validity will be substantiated through a sequence of results developed in algebro-geometric language in this section (see Theorem 3), although it may also be derived from the result of Ref.~\cite{BDMSST99} when viewed in the context of nUPBs, as in \cite{Pittenger,Skowronek11}.

%We begin by recalling several standard results (Claims \ref{claim1} and \ref{claim2}), reformulated in algebro-geometric language, which serve as building blocks for the forthcoming constructions. We proceed by formulating a statement whose validity will be substantiated through a sequence of results developed in this section (see Theorem~\ref{theorem-nqubit}).

\begin{manualclaim}{1}\label{claim1}
The following set is an nUPB for an $n$-qubit system
\begin{equation}\label{nUPB-qubit}
U_{2^n}=\{\otimes^n|q_l\rangle\}_{l=1}^{n+1}\,,
\end{equation}
where $|q_l\rangle=\alpha_l|0\rangle+\beta_l|1\rangle$ is the representation of a single-qubit pure state (being $\alpha_l,\beta_l\in\mathbbm{C}$ s.t. $|\alpha_l|^2+|\beta_l|^2=1$).
\end{manualclaim}

Geometrically, pure states of a single qubit can be represented as points on the surface of the Bloch sphere, which corresponds to the complex projective line $\mathbbm{C}\mathbbm{P}^1$ (hereafter, we denote $\mathbbm{C}\mathbbm{P}^d$ by $\mathbbm{P}^d$). Instead of using the homogeneous coordinates $[\alpha_l:\beta_l]$ for each point on the Bloch sphere (representing each qubit), one can opt for affine coordinates of the form $[1:x_l]$. In this representation, it is necessary to include the limit of $x_l$ to infinity, which corresponds to the point at infinity, represented by the state $|1\rangle\equiv[0 : 1]$. If we consider $n$ copies of a qubit state $|q_l\rangle$, i.e., the symmetric state $\otimes^n|q_l\rangle$, the affine coordinates of the resulting state are given by the image of the Veronese embedding \cite{Harris}:
\begin{equation}\label{Veronese-qubit}
\mathcal{V}_1^n\colon\mathbbm{P}^1\hookrightarrow\mathbbm{P}\big(\Sym^n(\mathbbm{C}^2)\big)\simeq\mathbbm{P}^{n}\,,
\end{equation}
that map each point $[1:x_l]\in\mathbbm{P}^1$ to $[1:x_l:x_l^2:\cdots:x_l^n]\in\mathbbm{P}^n$. Specifically, the $n$-fold tensor product of $|q_l\rangle$ is expressed as
\begin{equation}\label{n-qubit-copy}
\otimes^n|q_l\rangle=\sum_{k=0}^{n}\sum_{i_1+\cdots+i_n=k}x_l^{i_1+\cdots+i_n}|i_1\cdots i_n\rangle=\sum_{k=0}^{n}x_l^k|\mathrm{D}_n^k\rangle\,,
\end{equation}
where, for each $1\leq j\leq n$, $i_j\in\mathbbm{Z}_2$, and
\begin{equation}\label{Dicke-State}
|\mathrm{D}_n^k\rangle\coloneqq\sum_{\substack{i_1+\cdots+i_n=k \\ i_j\in\mathbbm{Z}_2}}|i_1\cdots i_n\rangle\,,
\end{equation}
are the unnormalized version of the well-known $n$-qubit Dicke states with $k$ excitations \cite{Dicke}. The set of Dicke states forms a basis for the symmetric subspace of the $n$-qubit Hilbert space. According to the binomial theorem, the normalization factor of each Dicke state $|\mathrm{D}_n^k\rangle$ is $\binom{n}{k}^{-1/2}$, where
\begin{equation}\label{binomial}
\binom{n}{k}\coloneqq\frac{n!}{k!(n-k)!}\,.
\end{equation}

\begin{manualremark}{2}\label{rmk-nqubit}
The Dicke states $\{|\mathrm{D}_n^k\rangle\}_{k=1}^{n-1}$ are GME because their $\ell$-multiranks are all strictly greater than one.
\end{manualremark}

\begin{manuallemma}{1}\label{lem-nqubit}
The subspace spanned by the GME Dicke states $\{|\mathrm{D}_n^k\rangle\}_{k=1}^{n-1}$ is a GES of dimension $n-1$.
\end{manuallemma}
{\it Proof.}
Consider the space spanned by all the Dicke states $\{|\mathrm{D}_n^k\rangle\}_{k=0}^n$, that is
\begin{equation}
\Sym^n(\mathbbm{C}^2)=\Span\{|\mathrm{D}_n^k\rangle\mid 0\leq k \leq n\}=\{\sum_{k=0}^{n}z_k|\mathrm{D}_n^k\rangle\mid z_k\in\mathbbm{C}\}\,.
\end{equation}
The $i$-th catalecticant matrix of a point in this space is defined as
\begin{equation}\label{catalecticant}
\mathrm{Cat}_i\coloneqq\begin{pmatrix}
z_0 & z_1 & \cdots & z_i \\
z_1 & z_2 & \cdots & z_{i+1} \\
\vdots & \vdots & \ddots & \vdots \\
z_{n-i} & z_{n-i+1} & \cdots & z_{n}
\end{pmatrix},
\end{equation}
for $1\leq i\leq n-1$ \cite{Sylvester,Gharahi24}. The rank of the catalecticant matrix in Eq. \eqref{catalecticant} is less than one iff all of its $2\times2$ minors vanish. For instance, for $z_i=x^i$, the rank of the catalecticant matrix is one (cf. Eq. \eqref{n-qubit-copy}). The subspace spanned by the GME Dicke states can be described as follows:
\begin{equation}\label{GME-Dick-subspace}
\Sym^n(\mathbbm{C}^2)\setminus\Span\{|\mathrm{D}_n^0\rangle,|\mathrm{D}_n^n\rangle\}=\Span\{|\mathrm{D}_n^k\rangle\mid 1\leq k \leq n-1\}=\{\sum_{k=1}^{n-1}z_k|\mathrm{D}_n^k\rangle\mid z_k\in\mathbbm{C}\}\,.
\end{equation}
The $i$-th catalecticant matrix of a point in this subspace is defined by Eq. \eqref{catalecticant}, with the conditions $z_0=z_n=0$. As a result, its rank is at least two for any choice of $z_1,\ldots,z_{n-1}\in\mathbbm{C}$, except when all coefficients $z_k$ vanish simultaneously, in which case it corresponds to zero. For binary symmetric tensors, the catalecticant matrices represent all the possible flattenings that do not have repeated rows and columns \cite{Sylvester,IK99}. Therefore, all the $\ell$-multiranks of any state in the subspace defined in Eq. \eqref{GME-Dick-subspace} are at least two.
\qed

By applying some elementary algebra, such as the Gram-Schmidt process, we can rewrite the non-orthogonal set $U_{2^n}$ of Eq. \eqref{nUPB-qubit} as a set of pairwise orthogonal Dicke states. Therefore, in light of Remark \ref{rmk-nqubit} and Lemma \ref{lem-nqubit}, we can state the following proposition.

\begin{manualproposition}{1}\label{prop-nqubit}
The subspace spanned by the nUPB in Eq. \eqref{nUPB-qubit} can be represented as follows
\begin{equation}\label{U-qubit-GES}
\mathcal{U}_{2^n}=\Span\{|\mathrm{D}_n^j\rangle\mid0\leq j\leq n\}=\Span\{\otimes^n|0\rangle,\otimes^n|1\rangle\}\oplus\Span\{|\mathrm{D}_n^j\rangle\mid1\leq j \leq n-1\}\,.
\end{equation}
\end{manualproposition}

We are now ready to formulate the main result of this section.

\begin{theorem}\label{theorem-nqubit}
An $n$-qubit Hilbert space $\mathcal{H}_{2^n}=\otimes^n\mathbbm{C}^2$ can be decomposed as follows
\begin{equation}\label{decHnqubit}
\mathcal{H}_{2^n}=\Span\{\otimes^n|0\rangle,\otimes^n|1\rangle\}\oplus\mathrm{GES}_{n-1}\oplus\mathrm{CES}_{2^n-n-1}\,,
\end{equation}
where $\mathrm{GES}_{n-1}$ refers to an $(n-1)$-dimensional GES containing GME Dicke states as its basis, and $\mathrm{CES}_{2^n-n-1}$ denotes a CES of the maximal dimension $2^n-n-1$. 
\end{theorem}
{\it Proof.}
The proof follows from Proposition \ref{prop-nqubit} with regard to the first two terms on the r.h.s. of Eq.\eqref{decHnqubit}. To complete the proof, we have to demonstrate that there is no fully separable state in the orthocomplement subspace of $\mathcal{U}_{2^n}$. We proceed by contradiction. Assume that there exists a fully separable state in the orthocomplement subspace. Let this state be denoted by $|\vartheta_1\cdots\vartheta_n\rangle$, where each $|\vartheta_j\rangle=a_j|0\rangle+b_j|1\rangle$ with fixed coefficients $a_j, b_j \in \mathbbm{C}$, representing a general qubit. Consequently, for all $x\in\mathbbm{C}$, we have
\begin{equation}
\langle\vartheta_1\cdots\vartheta_n|\otimes^n q\rangle= \langle\vartheta_1\cdots\vartheta_n|\sum_{i_1\in\mathbbm{Z}_{2}}\cdots \sum_{i_n\in\mathbbm{Z}_{2}}x^{i_1}\cdots x^{i_n}|i_1\cdots i_n\rangle=\prod_{j=1}^{n}\sum_{i_j\in\mathbbm{Z}_{2}}x^{i_j}\langle\vartheta_j|i_j\rangle=\prod_{j=1}^{n}(a^{*}_j+b^{*}_jx)=0\,.
\end{equation}
On the other hand, the last term of the above equation has only a finite number of solutions for $x$. Indeed, the number of solutions for each $j$ is bounded by the local dimension minus one, which in this case is $2-1=1$. This contradicts the assumption that the equation holds for all $x\in\mathbbm{C}$. Thus, we conclude that there is no fully separable state in the orthocomplement subspace of	$\mathcal{U}_{2^n}$. This completes the proof.
\qed

Regarding the last term on the r.h.s of Eq.\eqref{decHnqubit}, we may note that each GME Dicke state in the set $\{|\mathrm{D}_n^k\rangle\}_{k=1}^{n-1}$ is a uniform superposition of all permutations of $|0\rangle^{\otimes{(n-k)}} \otimes |1\rangle^{\otimes{k}}$, which yields $\binom{n}{k}$ terms. Let $2 \leq t \leq \binom{n}{k}$. By selecting a superposition of the first $t$ terms of a GME Dicke state with the following coefficients, one can generate a set of pairwise orthonormal states that are orthogonal to the primitive GME Dicke state
\begin{equation}\label{qubit-algorithm-CES}
\begin{tabular}{lccccccc}
$(t=2)$&\hspace{1cm}&$1$&$-1$&&&&\\
$(t=3)$&\hspace{1cm}&$1$&$1$&$-2$&&&\\
$(t=4)$&\hspace{1cm}&$1$&$1$&$1$&$-3$&&\\
$~~~~~\vdots$&\hspace{1cm}&$\vdots$&&&$\ddots$&$\ddots$&\\
$(t=\binom{n}{k})$&\hspace{1cm}&$1$&$1$&$1$&$\cdots$&$1$&$-(\binom{n}{k}-1)$\,.
\end{tabular}
\end{equation}
The normalization factor for each state in this set is $(t^2-t)^{-1/2}$. For each $k\in\{1,\ldots,n-1\}$, we obtain a pairwise orthonormal set of size $\binom{n}{k}-1$. Clearly, $\sum_{k=1}^{n-1}(\binom{n}{k}-1)=2^n-n-1$. Therefore, by following this algorithm, we have constructed a pairwise orthonormal set of maximum size for the CES.

\begin{manualexample}{2}\label{exmp-nqubit}
For a four-qubit Hilbert space $\mathcal{H}_{2^4}=\mathbbm{C}^2\otimes\mathbbm{C}^2\otimes\mathbbm{C}^2\otimes\mathbbm{C}^2$, consider the following set as an nUPB
\[
U_{2^4}=\{\otimes^4(|0\rangle+x_l|1\rangle)\}_{l=1}^{5}\,.
\]
By orthogonalizing $U_{2^4}$ and applying Theorem \ref{theorem-nqubit}, we derive the following decomposition of the four-qubit Hilbert space:
\begin{align*}
\mathcal{H}_{2^4}=&\Span\{|0000\rangle,|1111\rangle\}\oplus\Span\{|\mathrm{D}_4^1\rangle,|\mathrm{D}_4^2\rangle,|\mathrm{D}_4^3\rangle\} \\
& \oplus\Span\{\frac{1}{\sqrt{2}}(|0001\rangle-|0010\rangle), \frac{1}{\sqrt{6}}(|0001\rangle+|0010\rangle-2|0100\rangle), \frac{1}{\sqrt{12}}(|0001\rangle+|0010\rangle+|0100\rangle-3|1000\rangle), \\
&\qquad\qquad \frac{1}{\sqrt{2}}(|0011\rangle-|0101\rangle), \frac{1}{\sqrt{6}}(|0011\rangle+|0101\rangle-2|0110\rangle), \frac{1}{\sqrt{12}}(|0011\rangle+|0101\rangle+|0110\rangle-3|1001\rangle), \\
&\qquad\qquad \frac{1}{\sqrt{20}}(|0011\rangle+|0101\rangle+|0110\rangle+|1001\rangle-4|1010\rangle), \\
&\qquad\qquad \frac{1}{\sqrt{30}}(|0011\rangle+|0101\rangle+|0110\rangle+|1001\rangle+|1010\rangle-5|1100\rangle), \\
&\qquad\qquad \frac{1}{\sqrt{2}}(|0111\rangle-|1011\rangle), \frac{1}{\sqrt{6}}(|0111\rangle+|1011\rangle-2|1101\rangle), \frac{1}{\sqrt{12}}(|0111\rangle+|1011\rangle+|1101\rangle-3|1110\rangle)\} \\
=&\Span\{|0000\rangle,|1111\rangle\}\oplus\mathrm{GES}_{3}\oplus\mathrm{CES}_{11}\,.
\end{align*}
\end{manualexample}

%%%%%%%%%%%%%%%%%%%%%%%%%%%%%%%%%%%%%%%%%%%%%%%%%

\section{Multiqudit case}\label{Sec.iv}

In this section, we generalize the procedure outlined in the previous section to an $n$-qudit system whose associated Hilbert space is given by \mbox{$\mathcal{H}_{d^n}=\otimes^n\mathbbm{C}^{d}$}.

%%%%%%%%%%%%%%%%%%%%%%%%%%%%%%%%
%%%%%%%%%%%%%%%%%%%%%%%%%%%%%%%%

\subsection{Veronese embedding}\label{n-qudit-Veronese}

Borrowing the procedure used in the multiqubit case, we consider the Veronese embedding, which corresponds to the symmetric subspace $\Sym^n(\mathbbm{C}^d)$ of the $n$-qudit Hilbert space $\mathcal{H}_{d^n}$. In fact, multiqudit symmetric fully separable states possess the structure of a Veronese variety. The Veronese embedding is given by
\begin{equation}\label{Veronese-qudit}
\mathcal{V}_{d-1}^n\colon\mathbbm{P}^{d-1}\hookrightarrow\mathbbm{P}\big(\Sym^n(\mathbbm{C}^d)\big)\simeq\mathbbm{P}^{m-1}\,,
\end{equation}
where $m=\binom{n+d-1}{d-1}$ denotes the dimension of the symmetric subspace and is used throughout this section.

We now present an assertion that will be validated by a series of results in this section (see Theorem~\ref{theorem-nqudit-veronese}).

\begin{manualclaim}{2}\label{claim2}
The following set is an nUPB for an $n$-qudit system
\begin{equation}\label{nUPB-qudit-2}
U_{d^n}^{m}=\{\otimes^n|\varphi_l\rangle\}_{l=1}^{m}\,,
\end{equation}
where $|\varphi_l\rangle=\sum_{i\in\mathbbm{Z}_d}x_{i,l}|i\rangle$, with $x_{i,l}\in\mathbbm{C}$ s.t. $\sum_{i=0}^{d-1}|x_{i,l}|^2=1$, is a qudit pure state of dimension $d$.
\end{manualclaim}

The coordinates of $n$-qudit states in Eq. \eqref{nUPB-qudit-2} can be obtained from the image of the Veronese embedding defined in Eq. \eqref{Veronese-qudit}, as follows
\begin{equation}\label{qudit-nUPB-to-gen-Dicke}
\otimes^{n}|\varphi_l\rangle=\mathop{\otimes}\limits_{j=1}^{n}\sum_{i_j\in\mathbbm{Z}_d}x_{i_j,l}|i_j\rangle=\sum_{\substack{k_0+\cdots+k_{d-1}=n \\ k_0,\ldots,k_{d-1}\geq0}}x_{0,l}^{k_0}\cdots x_{d-1,l}^{k_{d-1}}~|\mathrm{D}_n^{\bk}\rangle\,,
\end{equation}
where $\bk=\{(k_0,\ldots,k_{d-1})\mid \sum_{i\in\mathbbm{Z}_d}k_i=n,~k_i\geq0\}$, and 
\begin{equation}\label{gen-Dicke}
|\mathrm{D}_n^{\bk}\rangle\coloneqq\sum_{\mathfrak{p}\in\mathfrak{P}_{\mathrm{M}(\bk)}}\mathfrak{p}\{\mathop{\otimes}\limits_{i\in\mathbbm{Z}_d}|i\rangle^{\otimes k_i}\}\,,
\end{equation}
with $\mathfrak{p}$ denoting elements of the permutation group $\mathfrak{P}_{\mathrm{M}(\bk)}$ of the multiset $\mathrm{M}(\bk)=\{0^{k_0},\ldots,(d-1)^{k_{d-1}}\}$, are the $n$-qudit Dicke states (or generalized Dicke states) characterized by the excitation vector (or the occupation number vector) $\bk$. According to the multinomial theorem, the normalization factor of each generalized Dicke state $|\mathrm{D}_n^{\bk}\rangle$ is given by $\binom{n}{\bk}^{-1/2}$, where
\begin{equation}\label{multinomial}
\binom{n}{\bk}\coloneqq\frac{n!}{\prod_{i\in\mathbbm{Z}_d}k_i!}\,.
\end{equation}

\begin{manualremark}{3}\label{rmk-nqudit-veronese}
The generalized Dicke states $|\mathrm{D}_n^{\bk}\rangle$ with the excitation vector $\bk$ containing components $0\leq~k_i<n$ for all $0\leq~i\leq~d-1$ are GME since their $\ell$-multiranks are all strictly greater than one. In contrast, the generalized Dicke states $|\mathrm{D}_n^{\bk}\rangle$ with the excitation vector containing a single component $k_i=n$ are fully separable states, that is, $|\mathrm{D}_n^{\pi\{(n,0,\ldots,0)\}}\rangle=|i\cdots i\rangle$, where $\pi$ denotes elements of the permutation group of the multiset $\{n,0^{d-1}\}$.
\end{manualremark}

\begin{manuallemma}{2}\label{lem-nqudit-veronese}
The subspace spanned by the GME generalized Dicke states $\{|\mathrm{D}_n^{\bk}\rangle\mid\bk\neq\pi\{(n,0,\ldots,0)\}\}$ is a GES of dimension $m-d$.
\end{manuallemma}
{\it Proof.}
Let us denote the set of all generalized Dicke states in Eq. \eqref{gen-Dicke} by introducing a free index: $\{|\mathrm{D}_n^{\bk_{l}}\rangle\}_{l=1}^m$. Additionally, we assume that the set of excitation vectors $\{\bk_{l}=(k_0,\ldots,k_{d-1})_{l}\mid\sum_{i\in\mathbbm{Z}_d}k_i=n,~k_i\geq0\}_{l=1}^{m}$ is ordered such that the concatenated components $k_0\cdots k_{d-1}$ form a sequence of decreasing values as $l$ increases. The subspace spanned by all generalized Dicke states is then given by
\begin{equation}
\Sym^n(\mathbbm{C}^d)=\Span\{|\mathrm{D}_n^{\bk_{l}}\rangle\mid 1\leq l\leq m\}=\{\sum_{l=1}^{m}z_{l}|\mathrm{D}_n^{\bk_{l}}\rangle\mid z_{l}\in\mathbbm{C}\}\,.
\end{equation}
For any bipartition, the flattening of a given state $|\psi\rangle\in\Sym^n(\mathbbm{C}^d)$ reads
\begin{equation}\label{flattening-gen-Dicke}
\mathcal{M}_I[\psi]=\begin{pmatrix}
z_1  & \ast & \cdots & \ast & \ast \\
\ast & \ast & \cdots & \ast & \ast \\
\vdots & \vdots & \ddots & \vdots & \vdots \\
\ast & \ast & \cdots & \ast & \ast \\
\ast & \ast & \cdots & \ast & z_m
\end{pmatrix}.
\end{equation}
The rank of the matrix $\mathcal{M}_{I}$ is less than one iff all of its $2\times2$ minors vanish. This occurs exclusively when the coefficients $z_{l}$ either take the specific form given in Eq. \eqref{qudit-nUPB-to-gen-Dicke}, representing a fully separable state, or when all of them vanish simultaneously, corresponding to a state with a value of zero. This follows from the fact that generalized Dicke states are uniform superpositions of certain canonical basis elements, specifically chosen to preserve symmetry under all permutations of subsystems (cf. Eq. \eqref{catalecticant}). Moreover, different generalized Dicke states do not share a common canonical basis, reinforcing their structural distinction. On the other hand, the subspace spanned by all the GME generalized Dicke states is given by
\begin{equation}\label{Space-GME-gen-Dicke}
\Span\{|\mathrm{D}_n^{\bk}\rangle\mid\bk\neq\pi\{(n,0,\ldots,0)\}\}\,.
\end{equation}
Thus, for any state in the subspace of Eq. \eqref{Space-GME-gen-Dicke}, all possible flattenings resemble Eq. \eqref{flattening-gen-Dicke}, but with $z_1=z_m=0$, along with $d-2$ additional zero matrix elements. Consequently, setting $z_1=z_m=0$ ensures that all $\ell$-multiranks are at least two for any choice of $\{z_{l}\}_{l=2}^{m-1}\in\mathbbm{C}$, except when all coefficients $z_l$ vanish simultaneously, in which case it corresponds to zero. This concludes the proof.
\qed

Applying the Gram-Schmidt process, we can transform the nonorthogonal set $U_{d^n}^{m}$ from Eq. \eqref{nUPB-qudit-2} into a pairwise orthogonal set of generalized Dicke states. Consequently, leveraging Remark \ref{rmk-nqudit-veronese} and Lemma \ref{lem-nqudit-veronese}, we establish the following proposition.

\begin{manualproposition}{2}\label{prop-nqudit-veronese}
The subspace spanned by the nUPB in Eq. \eqref{nUPB-qudit-2} can be represented as follows
\begin{equation}\label{Um-qudit-GES}
\mathcal{U}_{d^n}^{m}=\Span\{|i\rangle^{\otimes n}\mid i\in\mathbbm{Z}_d\}\oplus\Span\{|\mathrm{D}_n^{\bk}\rangle\mid\bk\neq\pi\{(n,0,\ldots,0)\}\}\,.
\end{equation}
\end{manualproposition}

Next, we highlight the core result of this subsection.

\begin{theorem}\label{theorem-nqudit-veronese}
An $n$-qudit Hilbert space $\mathcal{H}_{d^n}=\otimes_{j=1}^n\mathbbm{C}^{d}$ can be decomposed as follows
\begin{equation}\label{decHnqudit-2}
\mathcal{H}_{d^n}=\Span\{|i\rangle^{\otimes n}\mid i\in\mathbbm{Z}_d\}\oplus\mathrm{GES}_{m-d}\oplus\mathrm{CES}_{d^n-m}\,,
\end{equation}
where $\mathrm{CES}_{d^n-m}$ and $\mathrm{GES}_{m-d}$ denote an $(d^n-m)$-dimensional CES and an $(m-d)$-dimensional GES containing GME generalized Dicke states as its basis, respectively.
\end{theorem}
{\it Proof.}
The proof follows from Proposition \ref{prop-nqudit-veronese} for what concerns the first two terms on the r.h.s. of Eq.\eqref{decHnqudit-2}. To complete the proof, we have to demonstrate that there is no fully separable state in the orthocomplement subspace of $\mathcal{U}_{d^n}^{m}$. We proceed by contradiction. Assume that there exists a fully separable state in the orthocomplement subspace. Let this state be denoted by $|\vartheta_1\cdots\vartheta_n\rangle$, where each $|\vartheta_j\rangle=\sum_{i\in\mathbbm{Z}_{d_j}}a_{i,j}|i\rangle$ with fixed coefficients $a_{i,j}\in\mathbbm{C}$, representing a general qudit in $\mathbbm{C}^{d}$. Consequently, for all $\{x_i\in\mathbbm{C}\}_{i=0}^{d-1}$, we have
\begin{equation}
\langle\vartheta_1\cdots\vartheta_n|\otimes^n\varphi\rangle= \langle\vartheta_1\cdots\vartheta_n|\sum_{i_1\in\mathbbm{Z}_{d}}\cdots \sum_{i_n\in\mathbbm{Z}_{d}}x_{i_1}\cdots x_{i_n}|i_1\cdots i_n\rangle=\prod_{j=1}^{n}\sum_{i_j\in\mathbbm{Z}_{d}}x_{i_j}\langle\vartheta_j|i_j\rangle=\prod_{j=1}^{n}\sum_{i_j\in\mathbbm{Z}_{d}}a^{*}_{i,j}x_{i_j}=0\,.
\end{equation}
On the other hand, the last term of the above equation has only a finite number of solutions for $\{x_i\}_{i=0}^{d-1}$. Indeed, the number of solutions for each $j$ is bounded by the local dimension minus one, which in this case is $d-1$. This contradicts the assumption that the equation holds for all $x\in\mathbbm{C}$. Thus, we conclude that there is no fully separable state in the orthocomplement subspace of	$\mathcal{U}_{d^n}^{m}$. This completes the proof.
\qed

Concerning the last term on the r.h.s. of Eq.\eqref{decHnqudit-2}, we may note that each GME generalized Dicke state in the set $\{|\mathrm{D}_n^{\bk}\rangle\mid\bk\neq\pi\{(n,0,\ldots,0)\}\}$ contains $\binom{n}{\bk}$ terms. Let $2\leq t\leq\binom{n}{\bk}$. Selecting a superposition of the first $t$ terms of a GME generalized Dicke state with the following coefficients, one can generate a set of pairwise orthonormal states that are orthogonal to the primitive GME generalized Dicke state
\begin{equation}\label{qudit-V-algorithm-CES}
\begin{tabular}{lccccccc}
$(t=2)$&\hspace{1cm}&$1$&$-1$&&&&\\
$(t=3)$&\hspace{1cm}&$1$&$1$&$-2$&&&\\
$(t=4)$&\hspace{1cm}&$1$&$1$&$1$&$-3$&&\\
$~~~~~\vdots$&\hspace{1cm}&$\vdots$&&&$\ddots$&$\ddots$&\\
$(t=\binom{n}{\bk})$&\hspace{1cm}&$1$&$1$&$1$&$\cdots$&$1$&$-(\binom{n}{\bk}-1)$\,.
\end{tabular}
\end{equation}
The normalization factor for each state in this set is $(t^2-t)^{-1/2}$. For each $\bk\neq\pi\{(n,0,\ldots,0)\}$, one obtains a pairwise orthonormal set of size $\binom{n}{\bk}-1$. Since
\begin{equation}
\sum_{\substack{k_0+\cdots+k_{d-1}=n \\ 0\leq k_0,\ldots,k_{d-1}<n}}(\binom{n}{k_0,\ldots,k_{d-1}}-1)=d^n-m\,,
\end{equation}
this algorithm generates a pairwise orthonormal set of $(d^n-m)$-dimensional CES.

\begin{manualexample}{3}\label{exmp3}
For a three-qutrit Hilbert space $\mathcal{H}_{3^3}=\mathbbm{C}^3\otimes\mathbbm{C}^3\otimes\mathbbm{C}^3$, consider the following set as an nUPB
\[
U_{3^3}^{10}=\{\otimes^3(x_l|0\rangle+y_l|1\rangle+z_l|2\rangle)\}_{l=1}^{10}\,.
\]
By orthogonalizing $U_{3^3}^{10}$ and applying Theorem \ref{theorem-nqudit-veronese}, we derive the following decomposition of the three-qutrit Hilbert space:
\begin{align*}
\mathcal{H}_{3^3}=&\Span\{|000\rangle,|111\rangle,|222\rangle\}\oplus\Span\{|\mathrm{D}_3^{(2,1,0)}\rangle,|\mathrm{D}_3^{(2,0,1)}\rangle,|\mathrm{D}_3^{(1,2,0)}\rangle,|\mathrm{D}_3^{(1,1,1)}\rangle,|\mathrm{D}_3^{(1,0,2)}\rangle,|\mathrm{D}_3^{(0,2,1)}\rangle,|\mathrm{D}_3^{(0,1,2)}\rangle\}\\
& \oplus\Span\{\tfrac{1}{\sqrt{2}}(|001\rangle-|010\rangle), \tfrac{1}{\sqrt{6}}(|001\rangle+|010\rangle-2|100\rangle), \tfrac{1}{\sqrt{2}}(|002\rangle-|020\rangle), \tfrac{1}{\sqrt{6}}(|002\rangle+|020\rangle-2|200\rangle),\\
&\qquad\qquad \tfrac{1}{\sqrt{2}}(|011\rangle-|101\rangle), \tfrac{1}{\sqrt{6}}(|011\rangle+|101\rangle-2|110\rangle), \tfrac{1}{\sqrt{2}}(|022\rangle-|202\rangle), \tfrac{1}{\sqrt{6}}(|022\rangle+|202\rangle-2|220\rangle),\\
&\qquad\qquad \tfrac{1}{\sqrt{2}}(|112\rangle-|121\rangle), \tfrac{1}{\sqrt{6}}(|112\rangle+|121\rangle-2|211\rangle), \tfrac{1}{\sqrt{2}}(|122\rangle-|212\rangle), \tfrac{1}{\sqrt{6}}(|122\rangle+|212\rangle-2|221\rangle),\\
&\qquad\qquad \tfrac{1}{\sqrt{2}}(|012\rangle-|021\rangle), \tfrac{1}{\sqrt{6}}(|012\rangle+|021\rangle-2|102\rangle), \tfrac{1}{\sqrt{20}}(|012\rangle+|021\rangle+|102\rangle+|120\rangle-4|201\rangle), \\
&\qquad\qquad \tfrac{1}{\sqrt{12}}(|012\rangle+|021\rangle+|102\rangle-3|120\rangle), \tfrac{1}{\sqrt{30}}(|012\rangle+|021\rangle+|102\rangle+|120\rangle+|201\rangle-5|210\rangle)\} \\
=&\Span\{|000\rangle,|111\rangle,|222\rangle\}\oplus\mathrm{GES}_{7}\oplus\mathrm{CES}_{17}\,.
\end{align*}
\end{manualexample}

We now prove that the GES in Theorem \ref{theorem-nqudit-veronese}, and thus in Theorem \ref{theorem-nqubit}, is a symmetric GES of maximum dimension. The key observation is that within the symmetric subspace, the only biseparable states are in fact fully separable. Indeed, a biseparable state $|\psi\rangle = |\phi\rangle_{S} \otimes |\chi\rangle_{\bar{S}}$ can be globally symmetric only if it takes the form $|\psi\rangle = \otimes^n |\varphi\rangle$ for some single-qudit state $|\varphi\rangle$. This fact can be understood by considering the flattenings of symmetric states, which give rise to $\ell$-multiranks, symmetric $\binom{n}{\ell}$-tuples of ranks. However, it is important to note that the converse does not hold: a symmetric multirank does not necessarily imply that the underlying state is fully symmetric. We thus conclude the following.

\begin{manualproposition}{3}\label{prop-sym-CES=sym-GES}
Given that all non-product symmetric states are GME, it follows that any symmetric CES is, in essence, a symmetric GES.
\end{manualproposition}

\begin{manuallemma}{3}\label{lem-sym-CES}
Any subspace $S\subset\Sym^n(\mathbbm{C}^d)$ of dimension $m-d+1$ contains at least one nonzero product state.
\end{manuallemma}
{\it Proof.}
The subspace $S$ has a projectivization $\mathbbm{P}(S)$ of dimension $\dim{S}-1$. On the other hand, the dimension of the image of the Veronese embedding $\mathrm{Im}(\mathcal{V}_{d-1}^n)$ in Eq. \eqref{Veronese-qudit} is $d-1$. Since $\dim\mathbbm{P}(S)+\dim\mathrm{Im}(\mathcal{V}_{d-1}^n)=m-1\geq\dim\mathbbm{P}^{m-1}$, by applying Theorem 1.22 from Ref. \cite{Shafarevich}, we conclude that $\mathbbm{P}(S)\cap\mathrm{Im}(\mathcal{V}_{d-1}^n)\neq\emptyset$. This implies that the subspace $S$ contains at least one nonzero product state.
\qed

Therefore, Lemma \ref{lem-sym-CES} establishes that $m-d$ serves as an upper bound on the maximum dimension of a symmetric subspace that does not intersect with the Veronese variety; otherwise, this subspace contains at least one fully separable state. In other words, this gives the maximum dimension of a symmetric CES. Combined with Proposition \ref{prop-sym-CES=sym-GES}, this leads to the following theorem characterizing the maximum possible dimension of a symmetric GES.

\begin{theorem}\label{max-dim-sym-GES}
In the symmetric subspace $\Sym^n(\mathbbm{C}^d)$ of an $n$-qudit Hilbert space, the maximum dimension of a GES is $m-d$.
\end{theorem}
{\it Proof.}
Regarding Lemma \ref{lem-nqudit-veronese}, $m-d$ is a lower bound for the maximum dimension of a symmetric GES. Moreover, in conjunction with Lemma \ref{lem-sym-CES} and Proposition \ref{prop-sym-CES=sym-GES}, it follows that this lower bound is tight. This completes the proof.
\qed

%%%%%%%%%%%%%%%%%%%%%%%%%%%%%%%%
%%%%%%%%%%%%%%%%%%%%%%%%%%%%%%%%

\subsection{Modified Segre-Veronese embedding}\label{n-qudit-Segre-Veronese}

Although the procedure in the previous subsection provides a decomposition of the multiqudit Hilbert space similar to the qubit case (Theorem \ref{theorem-nqudit-veronese}), the resulting CES is not of maximum dimension. This occurs because the dimension of the symmetric subspace exceeds $n(d-1)+1$ when $d\geq3$. 

To overcome this issue, it is helpful to also consider the Segre embedding given by
\begin{equation}\label{Segre-qudit}
\Sigma_{\mathbf{d-1}}^n\colon\mathbbm{P}^{d-1}\times\cdots\times \mathbbm{P}^{d-1}\hookrightarrow\mathbbm{P}^{d^n-1}\,,
\end{equation}
where $\mathbf{d-1}=(d-1,\ldots,d-1)$. By comparing the Veronese and Segre embeddings, i.e., Eqs. \eqref{Veronese-qudit} and \eqref{Segre-qudit}, we observe that the image of the Veronese embedding can be derived from the image of the Segre embedding by considering points with identical coordinates, followed by a projection that merges the repeated coordinates. In fact, as we extend beyond qubits, the number of coordinate-defining variables increases. By imposing as many constraints on the coordinates as possible, we can effectively reduce the dimension of the nUPB. In the next example, we introduce these constraints on the coordinates.
 
\begin{manualexample}{4}\label{exmp4}
The Segre embedding for a two-qutrit system is given by
\begin{equation}\label{Segre-two-qutrit}
\Sigma_{(2,2)}^2\colon\mathbbm{P}^2\times\mathbbm{P}^2\hookrightarrow\mathbbm{P}(\mathbbm{C}^3\otimes\mathbbm{C}^3)\simeq\mathbbm{P}^8\,,
\end{equation}
which maps a pair of points $([1:x_1:x_2],[1:y_1:y_2])\in\mathbbm{P}^2\times\mathbbm{P}^2$ to their product $[1:x_1:x_2:y_1:y_2:x_1y_1:x_1y_2:x_2y_1:x_2y_2]\in\mathbbm{P}^8$. 
In contrast, the Veronese embedding of degree two in three variables is given by
\begin{equation}\label{Veronese-two-qutrit}
\mathcal{V}_2^2\colon\mathbbm{P}^2\hookrightarrow\mathbbm{P}\big(\Sym(\mathbbm{C}^3\otimes\mathbbm{C}^3)\big)\simeq\mathbbm{P}^5\,,
\end{equation}
which maps the point $[1:x_1:x_2]\in\mathbbm{P}^2$ to $[1:x_1:x_2:x_1^2:x_1x_2:x_2^2]\in\mathbbm{P}^5$. Here, the coordinates of the image are obtained as the symmetric self-product of the original point’s coordinates. By comparing the images of the embeddings in Eqs. \eqref{Segre-two-qutrit} and \eqref{Veronese-two-qutrit}, one can derive the coordinates of the Veronese embedding from the Segre embedding by substituting $y_1=x_1$ and $y_2=x_2$ followed by a projection that merges repeated coordinates. Additionally, we can substitute $x_2=x_1^2$, resulting in coordinates corresponding to the image of a Veronese embedding restricted to a conic, mapping the point $[1:x:x^2]$ to $[1:x:x^2:x^2:x^3:x^4]$. Note that $x^2$ appears twice in this expression. A further projection can be performed to retain only one instance of $x^2$. According to representation theory, the conic embedding defined by $[1:x:x^2]$ can be extended to $\mathbbm{P}^2$, and the resulting image remains smooth in $\mathbbm{P}^4$ \cite[Proposition 3.2, for the case $r=2$, $d=4$]{AF93}.
\end{manualexample}

In general, this procedure can be understood as the following map
\begin{equation}\label{Veronese-proj}
\mathbbm{P}\big(\Sym^{d-1}(\mathbbm{C}^2)\big)\hookrightarrow\mathbbm{P}\big(\Sym^{n(d-1)}(\mathbbm{C}^2)\big)\,.
\end{equation}
Indeed, the aforementioned map can be interpreted as a composition of two maps, as follows
\begin{equation}\label{composition}
\mathbbm{P}\big(\Sym^{d-1}(\mathbbm{C}^2)\big)\xhookrightarrow{\mathcal{V}_{d-1}^n}\mathbbm{P}\big(\Sym^n\big(\Sym^{d-1}(\mathbbm{C}^2)\big)\big)\xrightarrow{\Pi}\mathbbm{P}\big(\Sym^{n(d-1)}(\mathbbm{C}^2)\big)\,,
\end{equation}
where the first map is the $n$-th Veronese embedding restricted to the conic, and the second map, $\Pi$, is a projection onto the repeated coordinates. Specifically, the projection map $\Pi$ takes coordinates $x^i$ that appear multiple times and merges them into a single coordinate $x^i$. More precisely, the projection map $\Pi$ is defined as follows
\begin{equation}\label{ProjectionMap}
\Pi\colon[1:x:\cdots:x^i:\cdots:x^i:\cdots:x^{n(d-1)}]
\mapsto[1:x:\cdots:x^{i-1}:x^i:x^{i+1}:\cdots:x^{n(d-1)}]\,.
\end{equation}

Hence, at the coordinate level, we have the Veronese embedding restricted to the conic formed by monomials of a single variable, followed by a projection that merges repeated coordinates.

It is worth noting that $\mathbbm{P}\big(\Sym^{d-1}(\mathbbm{C}^2)\big)$ is the image of the Veronese embedding of degree $(d-1)$ in two variables (cf. Eq.\eqref{Veronese-qubit}). Accordingly, the composition in Eq.\eqref{composition} can be naturally extended to what we refer to as a double-Veronese embedding. This construction can be realized as a Segre–Veronese embedding (denoted by $\mathcal{SV}_{\mathbf{d-1}}^n$) on $n$ copies of $\mathbbm{P}^1$ with multidegree $\mathbf{d-1} = (d-1, \ldots, d-1)$, followed by symmetrization. The resulting embedding can be expressed diagrammatically as follows
\begin{equation}\label{double-Veronese-SV}
\begin{array}{c@{}c@{}c@{}c}
\mathbbm{P}^1 
& \xhookrightarrow{\mathcal{V}_1^{d-1}} \mathbbm{P}\big(\Sym^{d-1}(\mathbbm{C}^2)\big) \xhookrightarrow{\mathcal{V}_{d-1}^n}
& \mathbbm{P}\big(\Sym^n(\Sym^{d-1}(\mathbbm{C}^2))\big)
& \xrightarrow{\Pi} \mathbbm{P}\big(\Sym^{n(d-1)}(\mathbbm{C}^2)\big)\,.
\\[1ex]
\Big\downarrow \, \Delta && \Big\uparrow \, \Sym &
\\[1ex]
(\mathbbm{P}^1)^n
& \xhookrightarrow{\mathcal{SV}_{\mathbf{d-1}}^n}
& \mathbbm{P}\big(\otimes^n\Sym^{d-1}(\mathbbm{C}^2)\big)
& 
\end{array}
\end{equation}
Here, $\Delta$ denotes the diagonal map, and $\Sym$ represents the symmetrization map, explicitly given by
\begin{align}\label{Delta}
\Delta &\colon [1:x]\mapsto([1:x],\ldots,[1:x])\,, \\ \label{Sym-map}
\mathrm{Sym} &\colon \otimes^n V \to \Sym^{n}(V)\,.
\end{align}
Therefore, the double-Veronese embedding can be expressed as a symmetrized Segre–Veronese embedding, namely,
\begin{equation}\label{VoV-SymoSV}
\mathcal{V}_{d-1}^{n} \circ \mathcal{V}_{1}^{d-1} = \Sym\circ\, \mathcal{SV}_{\bf{d-1}}^{n}\circ\Delta\,.
\end{equation}

We now present the primary embedding of this section, defined as the composition of the symmetrized Segre–Veronese embedding with the projection map in Eq.~\eqref{ProjectionMap}, namely, $\Pi\circ\Sym\circ\,\mathcal{SV}_{\mathbf{d-1}}^{n}\circ\Delta$. This embedding is given by
\begin{equation}\label{Gamma}
\Gamma_{d-1}^n\colon \mathbbm{P}^1 \hookrightarrow \mathbbm{P}\big(\Sym^{n(d-1)}(\mathbbm{C}^2)\big)\simeq\mathbbm{P}^{n(d-1)}\,.
\end{equation}
This map explains why we can begin with $n$ copies of a qubit and end up with just one variable. This perspective will be useful for generalizing the procedure to multipartite systems, which will be discussed in Section \ref{sec.v}.

Finally, using the map $\Gamma_{d-1}^n$ in Eq. \eqref{Gamma}, we can generate an nUPB of the minimum size, and thus a CES of the maximum dimension for $n$-qudit systems.

An assertion is put forward below, the justification of which unfolds through a sequence of results in this section (see Theorem~\ref{theorem-nqudit-SegreVeronese}).

\begin{manualclaim}{3}\label{claim3}
The following set is an nUPB for an $n$-qudit system
\begin{equation}\label{nUPB-qudit}
U_{d^n}^{n(d-1)+1}=\{\otimes^n|\varphi_l\rangle\}_{l=1}^{n(d-1)+1}\,,
\end{equation}
where $|\varphi_l\rangle=\sum_{i=0}^{d-1}x_l^i|i\rangle$ is a qudit pure state of dimension $d$, obtained from the image of the Veronese embedding of degree $d-1$ applied to a qubit. Taking $n$ copies of a qubit, we obtain the $n$ copies of the desired qudit.
\end{manualclaim}

The coordinates of $n$-qudit states in Eq. \eqref{nUPB-qudit} can be determined by the image of the map $\Gamma_{d-1}^n$ defined in Eq. \eqref{Gamma}, which encapsulates their geometric structure, as follows
\begin{equation}
\otimes^{n}|\varphi_l\rangle=\sum_{k=0}^{n(d-1)}\sum_{i_1+\cdots+i_n=k}x_l^{i_1+\cdots+i_n}|i_1\cdots i_n\rangle=\sum_{k=0}^{n(d-1)}x_l^k|\mathrm{G}_n^k\rangle\,,
\end{equation}
where for each $1\leq j\leq n$, $i_j\in\mathbbm{Z}_d$, and
\begin{equation}\label{G-State}
|\mathrm{G}_n^k\rangle\coloneqq\sum_{\substack{i_1+\cdots+i_n=k \\ i_j\in\mathbbm{Z}_d}}|i_1\cdots i_n\rangle\,,
\end{equation}
mimic the Dicke states used in the qubit case.

Thanks to Lemma \ref{lemmaA1} in Appendix \ref{AppxA}, the normalization factor of each state $|\mathrm{G}_n^k\rangle$ is given by $({\binom{n}{k}}_{d-1})^{-1/2}$, where
\begin{equation}\label{S(n,k,d-1)}
{\binom{n}{k}}_{d-1}:=\sum_{l=0}^{\min(n,\lfloor\frac{k}{d}\rfloor)}(-1)^l\binom{n}{l}\binom{k-ld+n-1}{n-1}\,.
\end{equation}

\begin{manualremark}{4}\label{rmk-nqudit-SegreVeronese}
The set $\{|\mathrm{G}_n^k\rangle\}_{k=0}^{n(d-1)}$ is a basis for the symmetric space $\Sym^{n(d-1)}(\mathbbm{C}^2)$ which is a subspace of the symmetric subspace of the Hilbert space $\mathcal{H}_{d^n}$ (cf. Eqs. \eqref{Veronese-qudit} and \eqref{Veronese-proj}).
\end{manualremark}

\begin{manualremark}{5}\label{rmk-nqudit-SegreVeronese-2}
Since all the $\ell$-multiranks of the states $\{|\mathrm{G}_n^k\rangle\}_{k=1}^{n(d-1)-1}$ are strictly greater than one, these states are GME.
\end{manualremark}

\begin{manuallemma}{4}\label{lem-nqudit-SegreVeronese}
The subspace spanned by the GME $\mathrm{G}$ states $\{|\mathrm{G}_n^k\rangle\}_{k=1}^{n(d-1)-1}$ is a GES of dimension $n(d-1)-1$.
\end{manuallemma}
{\it Proof.}
Thanks to the Remark \ref{rmk-nqudit-SegreVeronese}, the proof follows a similar reasoning to that of Lemma \ref{lem-nqubit}.
\qed

With a Gram-Schmidt process, one can rewrite the non-orthogonal set $U_{d^n}^{n(d-1)+1}$ of Eq. \eqref{nUPB-qudit} as a pairwise orthogonal set of $\mathrm{G}$ states. Accordingly, thanks to Remark \ref{rmk-nqudit-SegreVeronese-2} and Lemma \ref{lem-nqudit-SegreVeronese}, we arrive at the following Proposition.

\begin{manualproposition}{4}\label{prop-nqudit-SegreVeronese}
The subspace spanned by the nUPB in Eq. \eqref{nUPB-qudit} can be represented as follows
\begin{equation}\label{U-qudit-GES}
\mathcal{U}_{d^n}^{n(d-1)+1}=\Span\{\otimes^n|0\rangle,\otimes^n|d-1\rangle\}\oplus\Span\{|\mathrm{G}_n^k\rangle\mid 1\leq k \leq n(d-1)-1\}\,.
\end{equation}
\end{manualproposition}

Now, we present the main result of this subsection.

\begin{theorem}\label{theorem-nqudit-SegreVeronese}
An $n$-qudit Hilbert space $\mathcal{H}_{d^n}=\otimes^n\mathbbm{C}^d$ can be decomposed as follows
\begin{equation}\label{decHnqudit}
\mathcal{H}_{d^n}=\Span\{\otimes^n|0\rangle,\otimes^n|d-1\rangle\}\oplus\mathrm{GES}_{n(d-1)-1}\oplus\mathrm{CES}_{d^n-n(d-1)-1}\,,
\end{equation}
where $\mathrm{GES}_{n(d-1)-1}$ denotes an $(n(d-1)-1)$-dimensional GES containing GME $\mathrm{G}$ states as its basis, and $\mathrm{CES}_{d^n-n(d-1)-1}$ represents a CES of the maximal dimension $d^n-n(d-1)-1$. 
\end{theorem}
{\it Proof.}
The proof follows from Proposition \ref{prop-nqudit-SegreVeronese} regarding the first two terms on the r.h.s. of Eq.\eqref{decHnqudit}. Now, we have to demonstrate that there is no fully separable state in the orthocomplement subspace of $\mathcal{U}_{d^n}^{n(d-1)+1}$. We proceed by contradiction. Assume that there exists a fully separable state in the orthocomplement subspace. Let this state be denoted by $|\vartheta_1\cdots\vartheta_n\rangle$, where each $|\vartheta_j\rangle=\sum_{i\in\mathbbm{Z}_d}a_{i,j}|i\rangle$ with fixed coefficients $a_{i,j}\in\mathbbm{C}$, representing a general qudit. Consequently, for all $x\in\mathbbm{C}$, we have
\begin{equation}
\langle\vartheta_1\cdots\vartheta_n|\otimes^n\varphi\rangle= \langle\vartheta_1\cdots\vartheta_n|\sum_{i_1\in\mathbbm{Z}_{d}}\cdots \sum_{i_n\in\mathbbm{Z}_{d}}x^{i_1}\cdots x^{i_n}|i_1\cdots i_n\rangle=\prod_{j=1}^{n}\sum_{i_j\in\mathbbm{Z}_{d}}x^{i_j}\langle\vartheta_j|i_j\rangle=\prod_{j=1}^{n}\sum_{i_j\in\mathbbm{Z}_{d}}a^{*}_{i,j}x^{i_j}=0\,.
\end{equation}
On the other hand, the last term of the above equation has only a finite number of solutions for $x$. Indeed, the number of solutions for each $j$ is bounded by the local dimension minus one that is $d-1$. This contradicts the assumption that the equation holds for all $x\in\mathbbm{C}$. Thus, we conclude that there is no fully separable state in the orthocomplement subspace of	$\mathcal{U}_{d^n}$. This completes the proof.
\qed

Regarding the last term on the r.h.s. of Eq.\eqref{decHnqudit}, we observe that each GME $\mathrm{G}$ state in the set $\{|\mathrm{G}_n^k\rangle\}_{k=1}^{n(d-1)-1}$ contains ${\binom{n}{k}}_{d-1}$ terms. Let $t$ be such that $2\leq t\leq{\binom{n}{k}}_{d-1}$. By choosing a superposition of the first $t$ terms of a GME $\mathrm{G}$ state with the following coefficients, one can construct a set of pairwise orthonormal states that are orthogonal to the primitive GME $\mathrm{G}$ state
\begin{equation}\label{qudit-SV-algorithm-CES}
\begin{tabular}{lccccccc}
$(t=2)$&\hspace{1cm}&$1$&$-1$&&&&\\
$(t=3)$&\hspace{1cm}&$1$&$1$&$-2$&&&\\
$(t=4)$&\hspace{1cm}&$1$&$1$&$1$&$-3$&&\\
$~~~~~\vdots$&\hspace{1cm}&$\vdots$&&&$\ddots$&$\ddots$&\\
$(t={\binom{n}{k}}_{d-1})$&\hspace{1cm}&$1$&$1$&$1$&$\cdots$&$1$&$-({\binom{n}{k}}_{d-1}-1)$\,.
\end{tabular}
\end{equation}
The normalization factor for each state in this set is $(t^2-t)^{-1/2}$. For each $k\in\{1,\ldots,n(d-1)-1\}$, we obtain a pairwise orthonormal set of size ${\binom{n}{k}}_{d-1}-1$. Since $\sum_{k=1}^{n(d-1)-1}({\binom{n}{k}}_{d-1}-1)=d^n-n(d-1)-1$, we have generated a pairwise orthonormal set of the maximum size for the CES.

\begin{manualexample}{5}\label{exmp5}
For a three-qutrit Hilbert space $\mathcal{H}_{3^3}=\mathbbm{C}^3\otimes\mathbbm{C}^3\otimes\mathbbm{C}^3$, consider the following set as an nUPB
\[
U_{3^3}^{7}=\{\otimes^3(|0\rangle+x_l|1\rangle+x_l^2|2\rangle)\}_{l=1}^{7}\,.
\]
Applying Theorem \ref{theorem-nqubit} after orthogonalizing $U_{3^3}^{7}$, we arrive at the following decomposition of the three-qutrit Hilbert space:
\begin{align*}
\mathcal{H}_{3^3}=&\Span\{|000\rangle,|222\rangle\}\oplus\Span\{|\mathrm{G}_3^1\rangle,|\mathrm{G}_3^2\rangle,|\mathrm{G}_3^3\rangle,|\mathrm{G}_3^4\rangle,|\mathrm{G}_3^5\rangle\} \\
& \oplus\Span\{\frac{1}{\sqrt{2}}(|001\rangle-|010\rangle), \frac{1}{\sqrt{6}}(|001\rangle+|010\rangle-2|100\rangle), \\
&\qquad\qquad \frac{1}{\sqrt{2}}(|011\rangle-|101\rangle), \frac{1}{\sqrt{6}}(|011\rangle+|101\rangle-2|110\rangle), \frac{1}{\sqrt{12}}(|011\rangle+|101\rangle+|110\rangle-3|002\rangle),  \\
&\qquad\qquad \tfrac{1}{\sqrt{20}}(|011\rangle+|101\rangle+|110\rangle+|002\rangle-4|020\rangle), \tfrac{1}{\sqrt{30}}(|011\rangle+|101\rangle+|110\rangle+|002\rangle+|020\rangle-5|200\rangle), \\
&\qquad\qquad \frac{1}{\sqrt{2}}(|012\rangle-|021\rangle), \frac{1}{\sqrt{6}}(|012\rangle+|021\rangle-2|102\rangle), \frac{1}{\sqrt{12}}(|012\rangle+|021\rangle+|102\rangle-3|120\rangle),  \\
&\qquad\qquad \tfrac{1}{\sqrt{20}}(|012\rangle+|021\rangle+|102\rangle+|120\rangle-4|201\rangle),\tfrac{1}{\sqrt{30}}(|012\rangle+|021\rangle+|102\rangle+|120\rangle+|201\rangle-5|210\rangle), \\
&\qquad\qquad \frac{1}{\sqrt{42}}(|012\rangle+|021\rangle+|102\rangle+|120\rangle+|201\rangle+|210\rangle-6|111\rangle), \\
&\qquad\qquad \frac{1}{\sqrt{2}}(|022\rangle-|202\rangle), \frac{1}{\sqrt{6}}(|022\rangle+|202\rangle-2|220\rangle), \frac{1}{\sqrt{12}}(|022\rangle+|202\rangle+|220\rangle-3|112\rangle),  \\
&\qquad\qquad \tfrac{1}{\sqrt{20}}(|022\rangle+|202\rangle+|220\rangle+|112\rangle-4|121\rangle), \tfrac{1}{\sqrt{30}}(|022\rangle+|202\rangle+|220\rangle+|112\rangle+|121\rangle-5|211\rangle), \\
&\qquad\qquad \frac{1}{\sqrt{2}}(|122\rangle-|212\rangle), \frac{1}{\sqrt{6}}(|122\rangle+|212\rangle-2|221\rangle)\} \\
=&\Span\{|000\rangle,|222\rangle\}\oplus\mathrm{GES}_{5}\oplus\mathrm{CES}_{20}\,.
\end{align*}
\end{manualexample}

%%%%%%%%%%%%%%%%%%%%%%%%%%%%%%%%
%%%%%%%%%%%%%%%%%%%%%%%%%%%%%%%%

\subsection{From Veronese to modified Segre-Veronese embedding}\label{n-qudit-VtoSV}

Considering the decompositions of the multiqubit Hilbert space (see Theorems \ref{theorem-nqudit-veronese} and \ref{theorem-nqudit-SegreVeronese}), obtained using the Veronese embedding and the modified Segre-Veronese embedding (the map $\Gamma_{d-1}^n$ in Eq. \eqref{Gamma}), a direct yet nontrivial question arises: What lies between the Veronese and the Segre-Veronese embeddings? In this section, we address this question by examining all possible constraints on the affine coordinates of the points in $\mathbbm{P}^{d-1}$.

In general, a point in $\mathbbm{P}^{d-1}$ has affine coordinates of the form $[1:x_1:x_2:x_3:x_4:\cdots:x_{d-1}]$. Building on Example \ref{exmp3}, we consider the substitutions $x_j=x_1^j$ for $2\leq j\leq d-1$, which impose constraints on the coordinates. So the number of possible substitutions, denoted by $\mathfrak{k}$, satisfies $1\leq\mathfrak{k}\leq d-2$. For each value of $\mathfrak{k}$, the affine coordinates of a point in $\mathbbm{P}^{d-1}$ take the following forms
\begin{align}\label{substitutions}
\begin{array}{cc}
(\mathfrak{k}=1) & [1:x_1:x_1^2:x_3:x_4:\cdots:x_{d-1}] \\
(\mathfrak{k}=2) & [1:x_1:x_1^2:x_1^3:x_4:\cdots:x_{d-1}] \\
\vdots & \vdots \\
(\mathfrak{k}=d-2)& ~~[1:x_1:x_1^2:x_1^3:x_1^4:\cdots:x_1^{d-1}]\,.
\end{array} 
\end{align}

After applying the Veronese embedding to a point in $\mathbbm{P}^{d-1}$ with $\mathfrak{k}$ substitutions as specified in Eq. \eqref{substitutions}, the resulting image exhibits repeated coordinates. To avoid this redundancy, we introduce the projection map $\mathcal{P}_{\mathfrak{k}}$, which identifies and merges these repeated coordinates into a single coordinate. In mathematical terms, the projection map $\mathcal{P}_{\mathfrak{k}}$ is defined as follows
\begin{equation}\label{ProjectionMap-gen}
\mathcal{P}_{\mathfrak{k}}\colon[1:x_1:\cdots:x_1^{a_1}\cdots x_{d-1}^{a_{d-1}}:\cdots:x_1^{a_1}\cdots x_{d-1}^{a_{d-1}}:\cdots]
\mapsto[1:x_1:\cdots:x_1^{a_1}\cdots x_{d-1}^{a_{d-1}}:\cdots]\,.
\end{equation}

The $n$-th Veronese embedding of points in $\mathbbm{P}^{d-1}$,  incorporating $\mathfrak{k}$ substitutions as specified in Eq. \eqref{substitutions}, followed by the projection map $\mathcal{P}_{\mathfrak{k}}$, can be defined as the composition
\begin{equation}\label{modified-Veronese}
\mathcal{P}_{\mathfrak{k}}\,\circ\, _{\mathfrak{k}}\mathcal{V}_{d-1}^n\colon\mathbbm{P}^{d-1}\hookrightarrow\mathbbm{P}^{\mathfrak{M}_{\mathfrak{k}}-1}\,,
\end{equation}
where $\mathfrak{M}_{\mathfrak{k}}$ is the number of distinct monomials in the expansion of
\begin{equation}\label{poly}
F=(1+x_1+x_1^2+\cdots+x_1^{\mathfrak{k}+1}+x_{\mathfrak{k}+2}+\cdots+x_{d-1})^n\,.
\end{equation}
Using the binomial theorem, the expansion in Eq. \eqref{poly} can be written as
\begin{equation}
F=\sum_{K=0}^{n}\binom{n}{K}A^{K}B^{n-K}\,,
\end{equation}
where $A=1+x_1+x_1^2+\cdots+x_1^{\mathfrak{k}+1}$ and $B=x_{\mathfrak{k}+2}+\cdots+x_{d-1}$.

\begin{manualexample}{6}\label{exmp6}
For the special cases involving $n$ copies of a qutrit and two copies of a qudit, we have the following:
\begin{enumerate}
\item
In the case of $\mathbbm{P}^2$, the only possible substitution is $x_2=x_1^2$, which leads to the following expression
\begin{equation}
\mathcal{P}_{1}\,\circ\, {_{1}\mathcal{V}_{2}^n}\colon\mathbbm{P}^2\hookrightarrow\mathbbm{P}^{2n}\,.
\end{equation}

\item
Consider a degree-two Veronese embedding of points in $\mathbbm{P}^{d-1}$, which incorporates $\mathfrak{k}$ substitutions as presented in Eq. \eqref{substitutions}, followed by the projection map $\mathcal{P}_{\mathfrak{k}}$, is given by
\begin{equation}
\mathcal{P}_{\mathfrak{k}}\,\circ\, _{\mathfrak{k}}\mathcal{V}_{d-1}^2\colon\mathbbm{P}^{d-1}\hookrightarrow\mathbbm{P}^{\binom{d+1}{2}-1-\sum_{i=1}^{\mathfrak{k}}i}\,.
\end{equation}
For $\mathfrak{k}=d-2$, it can be verified that this map coincides with the map $\Gamma_{d-1}^2$, as defined in Eq. \eqref{Gamma}.
\end{enumerate}
\end{manualexample}

\begin{manualremark}{6}\label{rmk-nqudit-VSV}
In the special cases of no substitution ($\mathfrak{k}=0$) and full substitution ($\mathfrak{k}=d-2$), the projection map $\mathcal{P}_{\mathfrak{k}}$ defined in Eq. \eqref{ProjectionMap-gen} coincides with the identity map $\mathrm{id}$ and the projection map $\Pi$ from Eq. \eqref{ProjectionMap}, respectively. As a result, for $\mathfrak{k}=0$, the map defined in Eq. \eqref{modified-Veronese} reduces to the Veronese map $\mathcal{V}_{d-1}^n$ in Eq. \eqref{Veronese-qudit}, and for $\mathfrak{k}=d-2$, it reduces to the map $\Gamma_{d-1}^n$ defined in Eq. \eqref{Gamma}.
\end{manualremark}

We now present a statement, the validity of which will be established through the forthcoming results in this section (see Theorem~\ref{theorem-nqudit-VSV}).

\begin{manualclaim}{4}\label{claim4}
The following set is an nUPB for an $n$-qudit system
\begin{equation}\label{nUPB-qudit-3}
U_{d^n}^{\mathfrak{M}_{\mathfrak{k}}}=\{\otimes^n|\varphi_l\rangle\}_{l=1}^{\mathfrak{M}_{\mathfrak{k}}}\,,
\end{equation}
where $|\varphi_l\rangle=\sum_{i=0}^{\mathfrak{k}+1}x_{1,l}^i|i\rangle+\sum_{i=\mathfrak{k}+2}^{d-1}x_{i,l}|i\rangle$ is a qudit pure state of dimension $d$.
\end{manualclaim}

Using the image of the modified Veronese embedding defined in Eq. \eqref{modified-Veronese}, the coordinates of $n$-qudit states in Eq. \eqref{nUPB-qudit-3} can be determined as follows
\begin{equation}\label{qudit-nUPB-to-F}
\otimes^{n}|\varphi_l\rangle=\sum_{K=0}^{n}\sum_{s=0}^{K(\mathfrak{k}+1)}\sum_{\substack{k_{\mathfrak{k}+2}+\cdots+k_{d-1}=n-K \\ k_{\mathfrak{k}+2},\ldots,k_{d-1}\geq0}}x_{1,l}^{s}x_{\mathfrak{k}+2,l}^{k_{\mathfrak{k}+2}}\cdots x_{d-1,l}^{k_{d-1}}~|\mathscr{G}_{K,n}^{s,\bs}\rangle\,,
\end{equation}
where $\bs=\{(k_{\mathfrak{k}+2},\ldots,k_{d-1})\mid \sum_{i=\mathfrak{k}+2}^{d-1}k_i=n-K,~k_i\geq0\}$, and
\begin{equation}\label{G-State-VSV}
|\mathscr{G}_{K,n}^{s,\bs}\rangle\coloneqq\sum_{\sigma\in\mathfrak{P}_{\mathrm{M}}}\sigma\{|\mathrm{G}_{K}^s\rangle\otimes|\mathrm{D}_{n-K}^{\bs}\rangle\}\,,
\end{equation}
with $\sigma$ denoting elements of the permutation group $\mathfrak{P}_{\mathrm{M}}$ of the multiset $\mathrm{M}=\{\gamma^K,\delta^{n-K}\}$. In each $\sigma$, the first $K$ occurrences of $\gamma$ are replaced by $i_1,\ldots,i_K$, preserving their order as given in the expansion of $|\mathrm{G}_K^s\rangle$, while the remaining $n-K$ occurrences of $\delta$ are replaced by $i_{K+1},\ldots,i_{n}$, preserving their order as given in the expansion of $|\mathrm{D}_{n-K}^{\bs}\rangle$. Additionally, based on Eqs. \eqref{G-State} and \eqref{gen-Dicke}, we have
\begin{equation}
|\mathrm{G}_{K}^s\rangle=\sum_{\substack{i_1+\cdots+i_{K}=s \\ i_j\in\mathbbm{Z}_{\mathfrak{k}+2}}}|i_1\cdots i_{K}\rangle\,,
\qquad \text{and} \qquad
|\mathrm{D}_{n-K}^{\bs}\rangle=\sum_{\mathfrak{q}\in\mathfrak{P}_{\mathrm{M}(\bs)}}\mathfrak{q}\{\mathop{\otimes}\limits_{i=\mathfrak{k}+2}^{d-1}|i\rangle^{\otimes k_i}\}\,,
\end{equation}
with $\mathfrak{q}$ denoting elements of the permutation group $\mathfrak{P}_{\mathrm{M}(\bs)}$ of the multiset $\mathrm{M}(\bs)=\{(\mathfrak{k}+2)^{k_{\mathfrak{k}+2}},\ldots,(d-1)^{k_{d-1}}\}$.

According to the binomial and multinomial theorems and Lemma \ref{lemmaA1}, the normalization factor of each $F$ state $|\mathscr{G}_{K,n}^{s,\bs}\rangle$ is given by $\mathcal{N}^{-1/2}$, where
\begin{equation}\label{normalization-G-VSV}
\mathcal{N}=\binom{n}{K}\binom{n-K}{\bs}{\binom{K}{s}}_{d-1}.
\end{equation}

\begin{manualremark}{7}\label{rmk-nqudit-VSV-2}
For the special cases $K=n$ and $K=0$, we get
\begin{equation}
|\mathscr{G}_{n,n}^{s,(\bf{0})}\rangle\equiv|\mathrm{G}_{n}^s\rangle\,,
\qquad\qquad \text{and} \qquad\qquad
|\mathscr{G}_{0,n}^{0,\bs}\rangle\equiv|\mathrm{D}_{n}^{\bs}\rangle\,,
\end{equation}
where $\bs=\{(k_{\mathfrak{k}+2},\ldots,k_{d-1})\mid \sum_{i=\mathfrak{k}+2}^{d-1}k_i=n,~k_i\geq0\}$ and $0\leq s \leq n(\mathfrak{k}+1)$.
\end{manualremark}

\begin{manualremark}{8}\label{rmk-nqudit-VSV-3}
The states $|\mathscr{G}_{K,n}^{s,\bs}\rangle$ are fully separable states either when $K=n$ and $s\in\{0,n(\mathfrak{k}+1)\}$, or when $K=0$ and $\bs=\tilde{\pi}(n,0,\ldots,0)$, where $\tilde{\pi}$ denotes elements of the permutation group of the multiset $\{n,0^{d-\mathfrak{k}-3}\}$; otherwise, they are GME, as all their $\ell$-multiranks are strictly greater than one.
\end{manualremark}

\begin{manuallemma}{5}\label{lem-nqudit-VSV}
The subspace spanned by the GME $\mathscr{G}$ states $\{|\mathscr{G}_{K,n}^{s,\bs}\rangle\mid K=n:1\leq s\leq n(\mathfrak{k}+1)-1,~K=0:\bs\neq\tilde{\pi}(n,0,\ldots,0)\}$ is a GES of dimension $\mathfrak{M}_{\mathfrak{k}}-d+\mathfrak{k}$.
\end{manuallemma}
{\it Proof.}
This lemma is a special case of Lemma \ref{lem-nqudit-veronese} for $1\leq\mathfrak{k}\leq{d-1}$. It reduces to Lemma \ref{lem-nqudit-veronese} and Lemma \ref{lem-nqudit-SegreVeronese} when $\mathfrak{k}=0$ and $\mathfrak{k}=d-2$, respectively.
\qed

Using the Gram-Schmidt process, the non-orthogonal set $U_{d^n}^{\mathfrak{M}_{\mathfrak{k}}}$ from Eq. \eqref{nUPB-qudit-3} can be transformed into a pairwise orthogonal set of $\mathscr{G}$ states. Consequently, by applying Remark \ref{rmk-nqudit-VSV-3} and Lemma \ref{lem-nqudit-VSV}, we obtain the following proposition.

\begin{manualproposition}{5}\label{prop-nqudit-VSV}
The subspace spanned by the nUPB in Eq. \eqref{nUPB-qudit-3} can be represented as follows
\begin{align}\label{gen-Um-qudit-GES}
\mathcal{U}_{d^n}^{\mathfrak{M}_{\mathfrak{k}}}=\Span\{\otimes^n|0\rangle,\otimes^n|\mathfrak{k}+1\rangle,\otimes^n|i+\mathfrak{k}+2\rangle \mid i\in\mathbbm{Z}_{d-\mathfrak{k}-2}\} \oplus\Span\left\{|\mathscr{G}_{K,n}^{s,\bs}\rangle \biggm|
\begin{aligned}
K=n\,:\,& 1\leq s\leq n(\mathfrak{k}+1)-1 \\
K=0\,:\,& \bs\neq\tilde{\pi}(n,0,\ldots,0)
\end{aligned}
\right\}.
\end{align}
\end{manualproposition}

We now proceed to the central theorem of this section.

\begin{theorem}\label{theorem-nqudit-VSV}
An $n$-qudit Hilbert space $\mathcal{H}_{d^n}=\otimes_{j=1}^n\mathbbm{C}^{d}$ can be decomposed as follows
\begin{equation}\label{decHnqudit-3}
\mathcal{H}_{d^n}=\Span\{\otimes^n|0\rangle,\otimes^n|\mathfrak{k}+1\rangle,\otimes^n|i+\mathfrak{k}+2\rangle \mid i\in\mathbbm{Z}_{d-\mathfrak{k}-2}\} \oplus\mathrm{GES}_{\mathfrak{M}_{\mathfrak{k}}-d+\mathfrak{k}}\oplus\mathrm{CES}_{d^n-\mathfrak{M}_{\mathfrak{k}}}\,,
\end{equation}
where $\mathrm{CES}_{d^n-\mathfrak{M}_{\mathfrak{k}}}$ and $\mathrm{GES}_{\mathfrak{M}_{\mathfrak{k}}-d+\mathfrak{k}}$ denote an $(d^n-\mathfrak{M}_{\mathfrak{k}})$-dimensional CES and an $(\mathfrak{M}_{\mathfrak{k}}-d+\mathfrak{k})$-dimensional GES containing GME $\mathscr{G}$ states as its basis, respectively.
\end{theorem}
{\it Proof.}
The proof follows from Proposition \ref{prop-nqudit-VSV} regarding the first two terms on the r.h.s. of Eq.\eqref{decHnqudit-3}. To complete the proof, we have to demonstrate that there is no fully separable state in the orthocomplement subspace of $\mathcal{U}_{d^n}^{\mathfrak{M}_{\mathfrak{k}}}$. We proceed by contradiction. Assume that there exists a fully separable state in the orthocomplement subspace. Let this state be denoted by $|\vartheta_1\cdots\vartheta_n\rangle$, where each $|\vartheta_j\rangle=\sum_{i\in\mathbbm{Z}_{d_j}}a_{i,j}|i\rangle$ with fixed coefficients $a_{i,j}\in\mathbbm{C}$, representing a general qudit in $\mathbbm{C}^{d}$. Moreover, let us rename the coefficients of the qudits we have used in Eq. \eqref{nUPB-qudit-3} as follows
\begin{equation}
|\varphi_l\rangle=\sum_{i=0}^{\mathfrak{k}+1}x_{1,l}^i|i\rangle+\sum_{i=\mathfrak{k}+2}^{d-1}x_{i,l}|i\rangle\equiv\sum_{i\in\mathbbm{Z}_d}y_{i,l}|i\rangle\,.
\end{equation}
Consequently, for all $\{y_i\in\mathbbm{C}\}_{i=0}^{d-1}$, we have
\begin{equation}
\langle\vartheta_1\cdots\vartheta_n|\otimes^n\varphi\rangle= \langle\vartheta_1\cdots\vartheta_n|\sum_{i_1\in\mathbbm{Z}_{d}}\cdots \sum_{i_n\in\mathbbm{Z}_{d}}y_{i_1}\cdots y_{i_n}|i_1\cdots i_n\rangle=\prod_{j=1}^{n}\sum_{i_j\in\mathbbm{Z}_{d}}y_{i_j}\langle\vartheta_j|i_j\rangle=\prod_{j=1}^{n}\sum_{i_j\in\mathbbm{Z}_{d}}a^{*}_{i,j}y_{i_j}=0\,.
\end{equation}
On the other hand, the last term of the above equation has only a finite number of solutions for $\{y_i\}_{i=0}^{d-1}$. Indeed, the number of solutions for each $j$ is bounded by the local dimension minus one, which in this case is $d-1$. This contradicts the assumption that the equation holds for all $y\in\mathbbm{C}$. Thus, we conclude that there is no fully separable state in the orthocomplement subspace of	$\mathcal{U}_{d^n}^{\mathfrak{M}_{\mathfrak{k}}}$. This concludes the proof.
\qed

Regarding the last term on the r.h.s. of Eq.\eqref{decHnqudit-3}, we may note that each GME $\mathscr{G}$ state belongs to the following set
\begin{equation}\label{GME-G-set-VSV}
\{|\mathscr{G}_{0,n}^{0,\bs}\rangle\mid\bs\neq\tilde{\pi}(n,0,\ldots,0)\}\cup\{|\mathscr{G}_{K,n}^{s,\bs}\rangle\mid 1\leq K\leq{n-1}\}\cup\{|\mathscr{G}_{n,n}^{s,(\bf{0})}\rangle\mid 1\leq s\leq n(\mathfrak{k}+1)-1\}\,,
\end{equation}
contains $\mathcal{N}$ terms given by Eq. \eqref{normalization-G-VSV}. Let $t$ be such that $2\leq{t}\leq\mathcal{N}$. By selecting a superposition of the first $t$ terms of a GME $\mathscr{G}$ state, with the following coefficients, we can construct a set of pairwise orthonormal states that are orthogonal to the primitive GME $\mathscr{G}$ state
\begin{equation}\label{qudit-VSV-algorithm-CES}
\begin{tabular}{lccccccc}
$(t=2)$&\hspace{1cm}&$1$&$-1$&&&&\\
$(t=3)$&\hspace{1cm}&$1$&$1$&$-2$&&&\\
$(t=4)$&\hspace{1cm}&$1$&$1$&$1$&$-3$&&\\
$~~~~~\vdots$&\hspace{1cm}&$\vdots$&&&$\ddots$&$\ddots$&\\
$(t=\mathcal{N})$&\hspace{1cm}&$1$&$1$&$1$&$\cdots$&$1$&$-(\mathcal{N}-1)$\,.
\end{tabular}
\end{equation}
The normalization factor for each state in this set is given by $(t^2-t)^{-1/2}$. So for each GME $\mathscr{G}$ state in the set in Eq. \eqref{GME-G-set-VSV}, one obtains a set of pairwise orthonormal states of size $\mathcal{N}-1$. Since
\begin{equation}
\sum_{\substack{k_{\mathfrak{k}+2}+\cdots+k_{d-1}=n \\ 0\leq k_{\mathfrak{k}+2},\ldots,k_{d-1}<n}}(\binom{n}{k_{\mathfrak{k}+2},\ldots,k_{d-1}}-1)+\sum_{K=1}^{n-1}\sum_{s=0}^{K(\mathfrak{k}+1)}\sum_{\substack{k_{\mathfrak{k}+2}+\cdots+k_{d-1}=n-K \\ k_{\mathfrak{k}+2},\ldots,k_{d-1}\geq0}}(\mathcal{N}-1)+\sum_{s=1}^{n(\mathfrak{k}+1)-1}({\binom{K}{s}}_{d-1}-1)=d^n-\mathfrak{M}_{\mathfrak{k}}\,,
\end{equation}
this algorithm generates a pairwise orthonormal set of $(d^n-\mathfrak{M}_{\mathfrak{k}})$-dimensional CES.

\begin{manualexample}{7}\label{exmp7}
In this example, we explore different decompositions of a three-ququart Hilbert space, $\mathcal{H}_{4^3}=\mathbbm{C}^4\otimes\mathbbm{C}^4\otimes\mathbbm{C}^4$. A point in $\mathbbm{P}^{3}$ has affine coordinates of the form $[1:x_1:x_2:x_3]$, and the third Veronese embedding of points in $\mathbbm{P}^{4}$ is given by
\begin{equation*}
\mathrm{id}\,\circ\, _{0}\mathcal{V}_{3}^{3}\,\equiv\,\mathcal{V}_{3}^{3}\colon\mathbbm{P}^{3}\hookrightarrow\mathbbm{P}^{19}\,.
\end{equation*}
Setting $x_1\equiv{x}$, the possible substitutions in the affine coordinates of the points in $\mathbbm{P}^{3}$, as specified in Eq. \eqref{substitutions}, are
\begin{align*}
\begin{array}{cc}
(\mathfrak{k}=1) & [1:x:x^2:x_3] \\
(\mathfrak{k}=2) & [1:x:x^2:x^3].
\end{array} 
\end{align*}
In accordance with Eq. \eqref{modified-Veronese}, the third Veronese embedding of points in $\mathbbm{P}^{3}$,  incorporating $1\leq\mathfrak{k}\leq{2}$ substitutions from the aforementioned equation, followed by the projection map $\mathcal{P}_{\mathfrak{k}}$ defined in Eq. \eqref{ProjectionMap-gen}, is given by
\begin{align*}
\mathcal{P}_{1}\,\circ\, _{1}\mathcal{V}_{3}^{3} & \colon\mathbbm{P}^{3}\hookrightarrow\mathbbm{P}^{15}\,, \\
\Pi\,\circ\, _{2}\mathcal{V}_{3}^{3}\,\equiv\,\Gamma_3^3 & \colon\mathbbm{P}^{3}\hookrightarrow\mathbbm{P}^{9}\,.
\end{align*}
Referring to Remark \ref{rmk-nqudit-VSV} and Theorems \ref{theorem-nqudit-veronese} and \ref{theorem-nqudit-SegreVeronese}, we obtain the following decompositions of the Hilbert space $\mathcal{H}_{4^3}$
\begin{align*}
& \mathcal{H}_{4^3}=\Span\{|000\rangle,|111\rangle,|222\rangle,|333\rangle\}\oplus\mathrm{GES}_{16}\oplus\mathrm{CES}_{44}\,, \\
& \mathcal{H}_{4^3}=\Span\{|000\rangle,|333\rangle\}\oplus\mathrm{GES}_{8}\oplus\mathrm{CES}_{54}\,,
\end{align*}
where the details follow similarly from Examples \ref{exmp3} and \ref{exmp5}. For $\mathfrak{k}=1$, consider the following set as an nUPB
\[
U_{4^3}^{16}=\{\otimes^3(|0\rangle+x_l|1\rangle+x_l^2|2\rangle+y_l|3\rangle)\}_{l=1}^{16}\,.
\]
Applying Theorem \ref{theorem-nqudit-VSV} after orthogonalizing $U_{4^3}^{16}$, we arrive at the following decomposition of the three-ququart Hilbert space:
\begin{align*}
\mathcal{H}_{4^3}=&\Span\{|000\rangle,|222\rangle,|333\rangle\} \\
& \oplus\Span\{|\mathrm{G}_3^1\rangle,|\mathrm{G}_3^2\rangle,|\mathrm{G}_3^3\rangle,|\mathrm{G}_3^4\rangle,|\mathrm{G}_3^5\rangle,|\mathscr{G}_{2,3}^{0,(1)}\rangle,|\mathscr{G}_{2,3}^{1,(1)}\rangle,|\mathscr{G}_{2,3}^{2,(1)}\rangle,|\mathscr{G}_{2,3}^{3,(1)}\rangle,|\mathscr{G}_{2,3}^{4,(1)}\rangle,|\mathscr{G}_{1,3}^{0,(2)}\rangle,|\mathscr{G}_{1,3}^{1,(2)}\rangle,|\mathscr{G}_{1,3}^{2,(2)}\rangle\} \\
& \oplus\Span\{\frac{1}{\sqrt{2}}(|001\rangle-|010\rangle), \frac{1}{\sqrt{6}}(|001\rangle+|010\rangle-2|100\rangle), \frac{1}{\sqrt{2}}(|011\rangle-|101\rangle), \frac{1}{\sqrt{6}}(|011\rangle+|101\rangle-2|110\rangle), \\
&~~\quad\qquad \frac{1}{\sqrt{12}}(|011\rangle+|101\rangle+|110\rangle-3|002\rangle), \tfrac{1}{\sqrt{20}}(|011\rangle+|101\rangle+|110\rangle+|002\rangle-4|020\rangle), \\
&~~\quad\qquad \tfrac{1}{\sqrt{30}}(|011\rangle+|101\rangle+|110\rangle+|002\rangle+|020\rangle-5|200\rangle), \\
&~~\quad\qquad \frac{1}{\sqrt{2}}(|012\rangle-|021\rangle), \frac{1}{\sqrt{6}}(|012\rangle+|021\rangle-2|102\rangle), \frac{1}{\sqrt{12}}(|012\rangle+|021\rangle+|102\rangle-3|120\rangle), \\
&~~\quad\qquad \tfrac{1}{\sqrt{20}}(|012\rangle+|021\rangle+|102\rangle+|120\rangle-4|201\rangle), \tfrac{1}{\sqrt{30}}(|012\rangle+|021\rangle+|102\rangle+|120\rangle+|201\rangle-5|210\rangle), \\
&~~\quad\qquad \tfrac{1}{\sqrt{42}}(|012\rangle+|021\rangle+|102\rangle+|120\rangle+|201\rangle+|210\rangle-6|111\rangle), \\
&~~\quad\qquad \frac{1}{\sqrt{2}}(|022\rangle-|202\rangle), \frac{1}{\sqrt{6}}(|022\rangle+|202\rangle-2|220\rangle), \frac{1}{\sqrt{12}}(|022\rangle+|202\rangle+|220\rangle-3|112\rangle), \\
&~~\quad\qquad \tfrac{1}{\sqrt{20}}(|022\rangle+|202\rangle+|220\rangle+|112\rangle-4|121\rangle), \tfrac{1}{\sqrt{30}}(|022\rangle+|202\rangle+|220\rangle+|112\rangle+|121\rangle-5|211\rangle), \\
&~~\quad\qquad \frac{1}{\sqrt{2}}(|122\rangle-|212\rangle), \frac{1}{\sqrt{6}}(|122\rangle+|212\rangle-2|221\rangle), \frac{1}{\sqrt{2}}(|003\rangle-|030\rangle), \frac{1}{\sqrt{6}}(|003\rangle+|030\rangle-2|300\rangle), \\
&~~\quad\qquad \frac{1}{\sqrt{2}}(|013\rangle-|031\rangle), \frac{1}{\sqrt{6}}(|013\rangle+|031\rangle-2|103\rangle), \frac{1}{\sqrt{12}}(|013\rangle+|031\rangle+|103\rangle-3|130\rangle), \\
&~~\quad\qquad \tfrac{1}{\sqrt{20}}(|013\rangle+|031\rangle+|103\rangle+|130\rangle-4|301\rangle), \tfrac{1}{\sqrt{30}}(|013\rangle+|031\rangle+|103\rangle+|130\rangle+|301\rangle-5|310\rangle), \\
&~~\quad\qquad \frac{1}{\sqrt{2}}(|023\rangle-|032\rangle), \frac{1}{\sqrt{6}}(|023\rangle+|032\rangle-2|203\rangle), \frac{1}{\sqrt{12}}(|023\rangle+|032\rangle+|203\rangle-3|230\rangle), \\
&~~\quad\qquad \tfrac{1}{\sqrt{20}}(|023\rangle+|032\rangle+|203\rangle+|230\rangle-4|302\rangle), \tfrac{1}{\sqrt{30}}(|023\rangle+|032\rangle+|203\rangle+|230\rangle+|302\rangle-5|320\rangle), \\
&~~\quad\qquad \tfrac{1}{\sqrt{42}}(|023\rangle+|032\rangle+|203\rangle+|230\rangle+|302\rangle+|320\rangle-6|113\rangle), \\
&~~\quad\qquad \tfrac{1}{\sqrt{56}}(|023\rangle+|032\rangle+|203\rangle+|230\rangle+|302\rangle+|320\rangle+|113\rangle-7|131\rangle), \\
&~~\quad\qquad \tfrac{1}{\sqrt{72}}(|023\rangle+|032\rangle+|203\rangle+|230\rangle+|302\rangle+|320\rangle+|113\rangle+|131\rangle-8|311\rangle), \\
&~~\quad\qquad \frac{1}{\sqrt{2}}(|123\rangle-|132\rangle), \frac{1}{\sqrt{6}}(|123\rangle+|132\rangle-2|213\rangle), \frac{1}{\sqrt{12}}(|123\rangle+|132\rangle+|213\rangle-3|231\rangle),  \\
&~~\quad\qquad \tfrac{1}{\sqrt{20}}(|123\rangle+|132\rangle+|213\rangle+|231\rangle-4|312\rangle), \tfrac{1}{\sqrt{30}}(|123\rangle+|132\rangle+|213\rangle+|231\rangle+|312\rangle-5|321\rangle), \\
&~~\quad\qquad  \frac{1}{\sqrt{2}}(|223\rangle-|232\rangle), \frac{1}{\sqrt{6}}(|223\rangle+|232\rangle-2|322\rangle), \frac{1}{\sqrt{2}}(|033\rangle-|303\rangle), \frac{1}{\sqrt{6}}(|033\rangle+|303\rangle-2|330\rangle), \\
&~~\quad\qquad \frac{1}{\sqrt{2}}(|133\rangle-|313\rangle), \frac{1}{\sqrt{6}}(|133\rangle+|313\rangle-2|331\rangle), \frac{1}{\sqrt{2}}(|233\rangle-|323\rangle), \frac{1}{\sqrt{6}}(|233\rangle+|323\rangle-2|332\rangle), \} \\
\qquad=&\Span\{|000\rangle,|222\rangle,|333\rangle\}\oplus\mathrm{GES}_{13}\oplus\mathrm{CES}_{48}\,.
\end{align*}
\end{manualexample}

%%%%%%%%%%%%%%%%%%%%%%%%%%%%%%%%%%%%%%%%%%%%%%%%%

\section{Multipartite case}\label{sec.v}

Finally, we explore the case of an $n$-partite system (a system consisting of $n$ qudits, each with a distinct finite dimension) whose corresponding Hilbert space is given by $\mathcal{H}_{\delta}=\otimes_{j=1}^n\mathbbm{C}^{d_j}$.

Denoting the dimension of the Hilbert space by $\textswab{D}\coloneqq\prod_{j=1}^{n}d_j$, and letting
$\textswab{S}\coloneqq\sum_{j=1}^{n}(d_j-1)$, the maximum dimension of the CES is then $\textswab{D}-\textswab{S}-1$.

Similar to the spirit of the multiqubit and multiqudit cases, we can start with $n$ copies of a qubit. The Segre-Veronese embedding of $n$ copies of $\mathbbm{P}^1$'s in multidegree ${\bf{d_\jmath-1}}=(d_1-1,\ldots,d_n-1)$ is given by
\begin{equation}\label{Segre-Veronese-gen}
\mathcal{SV}_{\bf{d_\jmath-1}}^n\colon(\mathbbm{P}^1)^n\hookrightarrow\mathbbm{P}\Big(\mathop{\otimes}\limits_{j=1}^{n}\Sym^{d_j-1}(\mathbbm{C}^2)\Big)\,.
\end{equation}
The image of this map is an algebraic variety known as the Segre-Veronese variety of multidegree ${\bf{d_\jmath-1}}$, which corresponds to the set of partially symmetric fully separable states in $\mathcal{H}_{\delta}$. There exists a canonical multiplication map $\mu$ from the tensor product of symmetric powers to the symmetric power of the sum
\begin{equation}
\mu\colon\mathop{\otimes}\limits_{j=1}^{n}\Sym^{d_j-1}(\mathbbm{C}^2)\to\Sym^{\textswab{S}}(\mathbbm{C}^2)\,,
\end{equation}
which is a (unique up to scalar) linear map defined by multiplying homogeneous polynomials $f_j(u,v)\in\Sym^{d_j-1}(\mathbbm{C}^2)$ of degree $d_j-1$, which correspond to symmetric tensors, resulting in a homogeneous polynomial of degree $\textswab{S}$.
This map is $\mathrm{SL}(2,\mathbbm{C})$-equivariant and thus induces an $\mathrm{SL}(2,\mathbbm{C})$-equivariant projective morphism
\begin{equation}\label{P-mu-multiplication}
\mathbbm{P}(\mu)\colon\mathbbm{P}\Big(\mathop{\otimes}\limits_{j=1}^{n}\Sym^{d_j-1}(\mathbbm{C}^2)\Big)\to\mathbbm{P}\big(\Sym^{\textswab{S}}(\mathbbm{C}^2)\big)\,.
\end{equation}
Therefore, the composition of the Segre–Veronese embedding (as defined in Eq.~\eqref{Segre-Veronese-gen}) with the canonical multiplication map (given in Eqs.~\eqref{P-mu-multiplication}) and the diagonal map (defined in Eq.~\eqref{Delta}), namely, $\mathbbm{P}(\mu)\circ\mathcal{SV}_{\bf{d\jmath-1}}^n\circ\Delta$, yields the desired map as follows
\begin{equation}\label{Gamma-gen}
\mathbf{\Gamma}_{\bf{d_\jmath-1}}^n \colon \mathbbm{P}^1 \hookrightarrow \mathbbm{P}\big(\Sym^{\textswab{S}}(\mathbbm{C}^2)\big)\simeq\mathbbm{P}^{\textswab{S}}\,.
\end{equation}

It is worth noting that, in general, the tensor product of symmetric tensors results in a partially symmetric tensor.

Ultimately, by using the map $\mathbf{\Gamma}_{\bf{d_\jmath-1}}^n$ in Eq. \eqref{Gamma-gen}, we can generate an nUPB of minimum size, and consequently, a CES of maximum dimension for multipartite systems.

In the following, we present an assertion whose validity will be supported by the upcoming results in this section (see Theorem~\ref{theorem-nqudit-VSV}).

\begin{manualclaim}{5}\label{claim5}
The following set is an nUPB for an $n$-partite system
\begin{equation}\label{nUPB-multipartite}
U_{\delta}=\{\otimes_{j=1}^{n}|\varphi_l^{j}\rangle\}_{l=1}^{\textswab{S}+1}\,,
\end{equation}
where each state in the set $\{|\varphi_l^{j}\rangle=\sum_{i=0}^{d_j-1}x_l^i|i\rangle\}_{j=1}^n$ is a qudit pure state of dimension $d_j$, obtained from the image of the Veronese embeddings of multidegree ${\bf{d_\jmath-1}}$ applied to a qubit. Taking $n$ copies of a qubit, we obtain the desired $n$-partite state $\otimes_{j=1}^{n}|\varphi_l^{j}\rangle$.
\end{manualclaim}

Based on the image of the map $\mathbf{\Gamma}_{\bf{d_\jmath-1}}^n$ defined in Eq. \eqref{Gamma-gen}, the coordinates of $n$-partite states in Eq. \eqref{nUPB-multipartite} can be determined as follows
\begin{equation}
\mathop{\otimes}\limits_{j=1}^{n}|\varphi_l^{j}\rangle=\sum_{k=0}^{\textswab{S}}\sum_{i_1+\cdots+i_n=k}x_l^{i_1+\cdots+i_n}|i_1\cdots i_n\rangle=\sum_{k=0}^{\textswab{S}}x_l^k|\textswab{G}_n^k\rangle\,,
\end{equation}
where for each $1\leq j\leq n$, $i_j\in\mathbbm{Z}_{d_j}$, and
\begin{equation}\label{G-State-gen}
|\textswab{G}_n^k\rangle\coloneqq\sum_{\substack{i_1+\cdots+i_n=k \\ i_j\in\mathbbm{Z}_{d_j}}}|i_1\cdots i_n\rangle\,,
\end{equation}
generalize the $\mathrm{G}$ states in Eq. \eqref{G-State} used in the multiqudit case.

Thanks to Lemma \ref{lemmaA2} in Appendix \ref{AppxA}, the normalization factor of each state $|\textswab{G}_n^k\rangle$ is $({\binom{n}{k}}_{\bf{d_\jmath-1}}
)^{-1/2}$, which  is given by
\begin{equation}\label{S(n,k,dj-1)}
{\binom{n}{k}}_{\bf{d_\jmath-1}}
\coloneqq\sum_{\textsc{s}\subseteq\{1,\ldots,n\}}(-1)^{|\textsc{s}|}\binom{k-(\sum_{j\in\textsc{s}}d_j)+n-1}{n-1}\,,
\end{equation}
where $\textsc{s}$ denotes all the possible subsets of $\{1,\ldots,n\}$ with cardinality $0\leq|\textsc{s}|\leq n$. It is worth noting that the binomial coefficient is included only if $k\geq\sum_{j\in\textsc{s}}d_j$; otherwise, the contribution of that subset $\textsc{s}$ is zero.

\begin{manualremark}{9}\label{rmk-npartite-SegreVeronese}
The set $\{|\textswab{G}_n^k\rangle\}_{k=0}^{\textswab{S}}$ is a basis for a partially symmetric subspace $\Sym^{\textswab{S}}(\mathbbm{C}^2)$ of the Hilbert space $\mathcal{H}_{\delta}$.
\end{manualremark}

\begin{manualremark}{10}\label{rmk-npartite-SegreVeronese-2}
The states $\{|\textswab{G}_n^k\rangle\}_{k=1}^{\textswab{S}-1}$ are GME, as all their $\ell$-multiranks exceed one.
\end{manualremark}

\begin{manuallemma}{6}\label{lem-npartite-SegreVeronese}
The subspace spanned by the GME $\textswab{G}$ states $\{|\textswab{G}_n^k\rangle\}_{k=1}^{\textswab{S}-1}$ is a GES of dimension $\textswab{S}-1$.
\end{manuallemma}
{\it Proof.}
By leveraging Remark \ref{rmk-npartite-SegreVeronese}, the proof proceeds in a manner analogous to that of Lemma \ref{lem-nqubit}.
\qed

Using a Gram-Schmidt process, one can rewrite the non-orthogonal set $U$ of Eq.\eqref{nUPB-multipartite} as a pairwise orthogonal set of $\textswab{G}$ states. Thus, based on Remark \ref{rmk-npartite-SegreVeronese-2} and Lemma \ref{lem-npartite-SegreVeronese}, we derive the following Proposition.

\begin{manualproposition}{6}\label{prop-npartite-SegreVeronese}
The subspace spanned by the nUPB in Eq. \eqref{nUPB-multipartite} can be represented as follows
\begin{equation}\label{U-multipartite-GES}
\mathcal{U}_{\delta}=\Span\{|\textswab{G}_n^k\rangle\mid0\leq k\leq\textswab{S}\}=\Span\{\otimes^{n}|0\rangle,\otimes_{j=1}^{n}|d_j-1\rangle\}\oplus\Span\{|\textswab{G}_n^k\rangle\mid 1\leq k\leq\textswab{S}-1\}\,.
\end{equation}
\end{manualproposition}

We can now state the key result of this subsection.

\begin{theorem}\label{theorem-npartite}
A multipartite Hilbert space $\mathcal{H}_{\delta}=\otimes_{j=1}^n\mathbbm{C}^{d_j}$ can be decomposed as follows
\begin{equation}\label{decH-multipartite}
\mathcal{H}_{\delta}=\Span\{\otimes^{n}|0\rangle,\otimes_{j=1}^{n}|d_j-1\rangle\}\oplus\mathrm{GES}_{\textswab{S}-1}\oplus\mathrm{CES}_{\textswab{D}-\textswab{S}-1}\,,
\end{equation}
where $\mathrm{GES}_{\textswab{S}-1}$ denotes an $(\textswab{S}-1)$-dimensional GES containing GME $\textswab{G}$-states as its basis, and $\mathrm{CES}_{\textswab{D}-\textswab{S}-1}$ refers to a CES of the maximal dimension $\textswab{D}-\textswab{S}-1$.
\end{theorem}
{\it Proof.}
The proof follows from Proposition \ref{prop-npartite-SegreVeronese} for what concerns the first two terms on the r.h.s. of Eq.\eqref{decH-multipartite}. To complete the proof, we have to demonstrate that there is no fully separable state in the orthocomplement subspace of $\mathcal{U}_{\delta}$. We proceed by contradiction. Assume that there exists a fully separable state in the orthocomplement subspace. Let this state be denoted by $|\vartheta_1\cdots\vartheta_n\rangle$, where each $|\vartheta_j\rangle=\sum_{i\in\mathbbm{Z}_{d_j}}a_{i,j}|i\rangle$ with fixed coefficients $a_{i,j}\in\mathbbm{C}$, representing a general qudit in $\mathbbm{C}^{d_j}$. Consequently, for all $x\in\mathbbm{C}$, we have
\begin{align}\nonumber
\langle\vartheta_1\cdots\vartheta_n|\varphi^1\cdots\varphi^n\rangle &= \langle\vartheta_1\cdots\vartheta_n|\sum_{i_1\in\mathbbm{Z}_{d_1}}\cdots \sum_{i_n\in\mathbbm{Z}_{d_n}}x^{i_1}\cdots x^{i_n}|i_1\cdots i_n\rangle=\prod_{j=1}^{n}\sum_{i_j\in\mathbbm{Z}_{d_j}}x^{i_j}\langle\vartheta_j|i_j\rangle \\
&=\prod_{j=1}^{n}\sum_{i_j\in\mathbbm{Z}_{d_j}}a^{*}_{i,j}x^{i_j}=0\,.
\end{align}
On the other hand, the last term of the above equation has only a finite number of solutions for $x$. Indeed, the number of solutions for each $j$ is bounded by the local dimension minus one, which in this case is $d_j-1$. This contradicts the assumption that the equation holds for all $x\in\mathbbm{C}$. Thus, we conclude that there is no fully separable state in the orthocomplement subspace of	$\mathcal{U}_{\delta}$. This completes the proof.
\qed

Concerning the last term on the r.h.s. of Eq.\eqref{decH-multipartite}, we may note that each GME $\textswab{G}$ state in the set $\{|\textswab{G}_n^k\rangle\}_{k=1}^{\textswab{S}-1}$ contains ${\binom{n}{k}}_{\bf{d_\jmath-1}}$ terms. Let $2\leq t\leq{\binom{n}{k}}_{\bf{d_\jmath-1}}$. Selecting a superposition of the first $t$ terms of a GME $\textswab{G}$ state with the following coefficients, one can create a set of pairwise orthonormal states that are orthogonal to the primitive GME $\textswab{G}$ state
\begin{equation}\label{gen-algorithm-CES}
\begin{tabular}{lccccccc}
$(t=2)$&\hspace{1cm}&$1$&$-1$&&&&\\
$(t=3)$&\hspace{1cm}&$1$&$1$&$-2$&&&\\
$(t=4)$&\hspace{1cm}&$1$&$1$&$1$&$-3$&&\\
$~~~~~\vdots$&\hspace{1cm}&$\vdots$&&&$\ddots$&$\ddots$&\\
$(t={\binom{n}{k}}_{\bf{d_\jmath-1}})$&\hspace{1cm}&$1$&$1$&$1$&$\cdots$&$1$&$-({\binom{n}{k}}_{\bf{d_\jmath-1}}-1)$\,.
\end{tabular}
\end{equation}
The normalization factor for each state in this set is $(t^2-t)^{-1/2}$. For each $k\in\{1,\ldots,\textswab{S}-1\}$, one obtains a pairwise orthonormal set of size ${\binom{n}{k}}_{\bf{d_\jmath-1}}-1$. Since $\sum_{k=1}^{\textswab{S}-1}({\binom{n}{k}}_{\bf{d_\jmath-1}}-1)=\textswab{D}-\textswab{S}-1$, this algorithm generates a pairwise orthonormal set of maximum size for the CES.

%%%%%%%%%%%%%%%%%%%%%%
%%%%%%%%%%%%%%%%%%%%%%

\begin{manualexample}{8}\label{exmp8}
For a qubit-qutrit-ququart Hilbert space $\mathcal{H}_{2\times3\times4}=\mathbbm{C}^2\otimes\mathbbm{C}^3\otimes\mathbbm{C}^4$, consider the following set as an nUPB
\[
U_{2\times3\times4}=\{(|0\rangle+x_l|1\rangle)\otimes(|0\rangle+x_l|1\rangle+x_l^2|2\rangle)\otimes(|0\rangle+x_l|1\rangle+x_l^2|2\rangle+x_l^3|3\rangle)\}_{l=1}^{7}\,.
\]
After performing orthogonalization on $U_{2\times3\times4}$ and applying Theorem \ref{theorem-npartite}, we obtain the following decomposition of the qubit-qutrit-ququart Hilbert space:
\begin{align*}
\mathcal{H}_{2\times3\times4}=&\Span\{|000\rangle,|123\rangle\}\oplus\Span\{|\textswab{G}_3^1\rangle,|\textswab{G}_3^2\rangle,|\textswab{G}_3^3\rangle,|\textswab{G}_3^4\rangle,|\textswab{G}_3^5\rangle\} \\
& \oplus\Span\{\frac{1}{\sqrt{2}}(|001\rangle-|010\rangle), \frac{1}{\sqrt{6}}(|001\rangle+|010\rangle-2|100\rangle), \frac{1}{\sqrt{2}}(|011\rangle-|101\rangle),  \\
& \qquad\qquad \frac{1}{\sqrt{6}}(|011\rangle+|101\rangle-2|110\rangle), \frac{1}{\sqrt{12}}(|011\rangle+|101\rangle+|110\rangle-3|002\rangle), \\
& \qquad\qquad \tfrac{1}{\sqrt{20}}(|011\rangle+|101\rangle+|110\rangle+|002\rangle-4|020\rangle), \frac{1}{\sqrt{2}}(|012\rangle-|021\rangle), \frac{1}{\sqrt{6}}(|012\rangle+|021\rangle-2|102\rangle), \\
& \qquad\qquad  \frac{1}{\sqrt{12}}(|012\rangle+|021\rangle+|102\rangle-3|120\rangle), \tfrac{1}{\sqrt{20}}(|012\rangle+|021\rangle+|102\rangle+|120\rangle-4|111\rangle),  \\
& \qquad\qquad \tfrac{1}{\sqrt{30}}(|012\rangle+|021\rangle+|102\rangle+|120\rangle+|111\rangle-5|003\rangle), \\
& \qquad\qquad \frac{1}{\sqrt{2}}(|013\rangle-|022\rangle), \frac{1}{\sqrt{6}}(|013\rangle+|022\rangle-2|103\rangle), \frac{1}{\sqrt{12}}(|013\rangle+|022\rangle+|103\rangle-3|112\rangle), \\
& \qquad\qquad \tfrac{1}{\sqrt{20}}(|013\rangle+|022\rangle+|103\rangle+|112\rangle-4|121\rangle), \frac{1}{\sqrt{2}}(|023\rangle-|113\rangle), \frac{1}{\sqrt{6}}(|023\rangle+|113\rangle-2|122\rangle) \} \\
=&\Span\{|000\rangle,|123\rangle\}\oplus\mathrm{GES}_{5}\oplus\mathrm{CES}_{17}\,.
\end{align*}
\end{manualexample}

\begin{manualremark}{11}\label{rmk-GME-algorithm}
All the $\ell$-multiranks of the states defined by the last-line coefficients in Eq. \eqref{gen-algorithm-CES}, as well as in Eqs. \eqref{qubit-algorithm-CES}, \eqref{qudit-V-algorithm-CES}, \eqref{qudit-SV-algorithm-CES}, and \eqref{qudit-VSV-algorithm-CES}, denoted as $|\textswab{F}_n^k\rangle$, are strictly greater than one, confirming that these states are GME. Furthermore, their span forms a GES with dimensions $\textswab{S}-1$, $n-1$, $m-d$, $n(d-1)-1$, and $\mathfrak{M}_{\mathfrak{k}}-d+\mathfrak{k}$, respectively.
\end{manualremark}

Using Remark \ref{rmk-GME-algorithm} and building on Theorem \ref{theorem-npartite}, which generalizes Theorems \ref{theorem-nqudit-SegreVeronese} and \ref{theorem-nqubit}, we derive the following result.

\begin{theorem}\label{theorem-GES-GES-CES}
A multipartite Hilbert space $\mathcal{H}_{\delta}=\otimes_{j=1}^n\mathbbm{C}^{d_j}$ can be decomposed as follows
\begin{equation}\label{decH-GES-GES-CES}
\mathcal{H}_{\delta}=\Span\{\otimes^{n}|0\rangle,\otimes_{j=1}^{n}|d_j-1\rangle\}\oplus\mathrm{GES}_{\textswab{S}-1}\oplus\mathrm{GES}_{\textswab{S}-1}\oplus\mathrm{CES}_{\textswab{D}-2\,\textswab{S}}\,,
\end{equation}
where $\mathrm{GES}_{\textswab{S}-1}$ denote $(\textswab{S}-1)$-dimensional GESs and $\mathrm{CES}_{\textswab{D}-2\,\textswab{S}}$ refers to a $(\textswab{D}-2\,\textswab{S})$-dimensional CES.
\end{theorem}
{\it Proof.}
The proof follows from Theorem \ref{theorem-npartite} and Remark \ref{rmk-GME-algorithm}.
\qed

\begin{manualexample}{9}\label{exmp9}
In Example \ref{exmp8}, the $\textswab{F}$ states $|\textswab{F}_n^k\rangle$ are given as follows
\begin{align*}
& |\textswab{F}_3^1\rangle=\frac{1}{\sqrt{6}}(|001\rangle+|010\rangle-2|100\rangle)\,, \\
& |\textswab{F}_3^2\rangle=\frac{1}{\sqrt{20}}(|011\rangle+|101\rangle+|110\rangle+|002\rangle-4|020\rangle)\,, \\
& |\textswab{F}_3^3\rangle=\frac{1}{\sqrt{30}}(|012\rangle+|021\rangle+|102\rangle+|120\rangle+|111\rangle-5|003\rangle)\,, \\
& |\textswab{F}_3^4\rangle=\frac{1}{\sqrt{20}}(|013\rangle+|022\rangle+|103\rangle+|112\rangle-4|121\rangle)\,, \\
& |\textswab{F}_3^5\rangle= \frac{1}{\sqrt{6}}(|023\rangle+|113\rangle-2|122\rangle)\,.
\end{align*}
Therefore, we obtain the following decomposition of the qubit-qutrit-ququart Hilbert space:
\begin{align*}
\mathcal{H}_{2\times3\times4}&=\Span\{|000\rangle,|123\rangle\}\oplus\{|\textswab{G}_3^1\rangle,|\textswab{G}_3^2\rangle,|\textswab{G}_3^3\rangle,|\textswab{G}_3^4\rangle,|\textswab{G}_3^5\rangle\}\oplus\Span\{|\textswab{F}_3^1\rangle,|\textswab{F}_3^2\rangle,|\textswab{F}_3^3\rangle,|\textswab{F}_3^4\rangle,|\textswab{F}_3^5\rangle\}\oplus\mathrm{CES}_{12}\\
&=\Span\{|000\rangle,|123\rangle\}\oplus\mathrm{GES}_{5}\oplus\mathrm{GES}_{5}\oplus\mathrm{CES}_{12}\,.
\end{align*}
\end{manualexample}

%%%%%%%%%%%%%%%%%%%%%%%%%%%%%%%%%%%%%%%%%%%%%%%%%

\section{Conclusions and outlook}\label{Sec.vi}

Inspired by principles from algebraic geometry, we have proposed a novel approach for constructing entangled subspaces within the Hilbert space of multipartite quantum systems by embedding an nUPB into an algebraic variety. In multipartite systems, imposing maximal constraints on the coordinates of the nUPB points effectively minimizes their dimensionality. To attain the smallest possible nUPB and, consequently, the largest possible CES, we employed a modified Veronese embedding—actually, a modified Segre-Veronese embedding—which imposes the maximum possible constraints on the coordinates of the nUPB. Furthermore, through an orthogonalization procedure such as the Gram-Schmidt process, we transformed the nUPB into a set of pairwise orthogonal states. Notably, due to the symmetry inherent in the nUPB, we observed that, after orthogonalization, a significant subset of these states could be converted into a GES. We then systematically investigated the transition from the standard Veronese embedding to the modified Segre–Veronese embedding by imposing various constraints on the affine coordinates. This procedure results in an increase of the CES dimension, while reducing that of the GES, thereby leading to multiple possible decompositions of the Hilbert space.

Our results demonstrate that algebraic geometry provides a rigorous and effective framework for identifying properties of entangled subspaces. Actually, previous findings on the construction of GES of any size \cite{Demianowicz22} can be naturally expressed in the language of algebraic geometry. In particular, such constructions appear to rely on a Segre embedding, where specific coefficients are assigned to the points. For instance, in the case of a multiqubit system, they use $[1:x^{2^{n-1}}]$ $\times$ $[1:x^{2^{n-2}}]$ $\times$ $\cdots$ $\times$ $[1:x]$ as the coefficients for the $n$ qubits, which are then mapped to $[1:x:x^2:\cdots:x^{2^n-1}]$. While the method in Ref. \cite{Demianowicz22} does not allow for the construction of a symmetric GES, our approach not only achieves this but also constructs a symmetric GES of maximum dimension. Given that our approach targets highly symmetric subspaces of the Hilbert space, we are confident that the characterization of entangled states, in a manner similar to Refs.~\cite{GMO20,GM21}, can be carried out effectively.

Motivated by Theorem \ref{theorem-GES-GES-CES}, which demonstrates that multiple  GESs can be distinguished, we present in Appendix \ref{AppxB} an alternative algorithm that not only provides a constructive procedure for generating the CESs described in Theorems \ref{theorem-nqubit}-\ref{theorem-nqudit-veronese} and \ref{theorem-nqudit-SegreVeronese}-\ref{theorem-npartite} but also enables the extraction of GESs from the corresponding CESs. This naturally raises an intriguing open question: Can an arbitrary Hilbert space be decomposed entirely into mutually orthogonal GESs (or CESs)? The answer is yes—in principle—such decompositions do exist. In fact, a random decomposition of the Hilbert space into subspaces of suitable dimensions will almost surely produce GESs, as guaranteed by results on typical entanglement in high-dimensional spaces \cite{WS08,HLW04}. What remains open, however, is how to obtain such decompositions in a controlled, explicit, and algebraically tractable manner. Appendix~\ref{AppxC} provides a partial step in this direction by presenting a fully explicit decomposition of the three-qubit Hilbert space into three mutually orthogonal GESs. Moreover, as suggested by Refs.~\cite{LPS06,MCF14}, another possible approach is to investigate the conditions under which superpositions of GME states remain GME, which may yield further constructive criteria.

Looking far afield it would be interesting to explore the connection between the Hilbert-space decompositions presented here and representation-theoretic methods, which may provide additional structural insights into entangled subspaces.

%%%%%%%%%%%%%%%%%%%%%%%%%%%%%%%%%%%%%%%%%%%%%%%%%

\section*{Acknowledgments}
The authors warmly thank Giorgio Ottaviani for his insightful discussions, valuable comments, and for bringing Ref. \cite{AF93} to our attention. M.~G. is also grateful to Seyed Javad Akhtarshenas for his careful proofreading and thoughtful feedback. M.~G. acknowledges financial support from ``MUR project PRIN 2022SW3RPY'', and from the European Union under ``ERC Advanced Grant TAtypic, Project No. 101142236''. S.~M. acknowledges financial support from the ``PNRR MUR project PE0000023-NQSTI''.
Views and opinions expressed are, however, those of the author(s) only and do not necessarily reflect those of the European Union or the European Research Council Executive Agency. Neither the European Union nor the granting authority can be held responsible for them.

%%%%%%%%%%%%%%%%%%%%%%%%%%%%%%%%%%%%%%%%%%%%%%%%%
\newpage
\appendix
%%%%%%%%%%%%%%%%%%%%%%%%%%%%%%%%%%%%%%%%%%%%%%%%%

\section{Normalization factor}\label{AppxA}

\begin{manuallemma}{A1}\label{lemmaA1}
The number of non-negative integer solutions to the equation $x_1+\cdots+x_n=k$ where $0\leq x_j \leq d-1$, for all $1\leq j\leq n$, is given by ${\binom{n}{k}}_{d-1}$ in Eq. \eqref{S(n,k,d-1)}.
\end{manuallemma}
{\it Proof.}
Each variable $x_j$ has the constraint $0\leq x_k \leq d-1$. So the generating function for $x_j$ is
\begin{equation}
\mathcal{G}(x_j)=1+x+x^2+\cdots+x^{d-1}=\frac{1-x^d}{1-x}\,.
\end{equation}
Since all variables $\{x_j\}_{j=1}^n$ are independent, the generating function for their sum is
\begin{equation}
\mathcal{G}(x_1,\ldots,x_n)=\left(\frac{1-x^d}{1-x}\right)^n\,.
\end{equation}
The number of solutions is equal to the coefficient of $x^k$ in the expansion of the generating function $\mathcal{G}(x_1,\ldots,x_n)$. By applying the binomial theorem and the power series expansion, we obtain
\begin{align}
(1-x^d)^n &=\sum_{l=0}^{n}(-1)^l\binom{n}{l}x^{ld}\,, \\ \label{a4}
\frac{1}{(1-x)^n} &=\sum_{m=0}^{\infty}\binom{m+n-1}{n-1}x^m\,.
\end{align}
By multiplying the two expansions and collecting the coefficient of $x^k$, we obtain the condition $m+ld=k$, or equivalently, $m=k-ld$ (noting that $k-ld\geq0$). This leads to
\begin{equation}
[x^k]\mathcal{G}(x_1,\ldots,x_n)=\sum_{l=0}^{\min(n,\lfloor\frac{k}{d}\rfloor)}(-1)^l\binom{n}{l}\binom{k-ld+n-1}{n-1}\,,
\end{equation}
where $[x^k]\mathcal{G}(x_1,\ldots,x_n)$ denotes the coefficient of $x^k$ in the expansion of the generating function $\mathcal{G}(x_1,\ldots,x_n)$. Note that the upper bound on the summation index $l$ is $\min(n,\lfloor\frac{k}{d}\rfloor)$, since we require $ld\leq k$ to assure $k-ld\geq0$, and additionally, $k\leq n$ as there are $n$ variables.
\qed

\begin{manuallemma}{A2}\label{lemmaA2}
The number of non-negative integer solutions to the equation $x_1+\cdots+x_n=k$ where $0\leq x_j \leq d_j-1$, for all $1\leq j\leq n$, is given by ${\binom{n}{k}}_{\bf{d_\jmath-1}}$ in Eq. \eqref{S(n,k,dj-1)}.
\end{manuallemma}
{\it Proof.}
Each variable $x_j$ has the constraint $0\leq x_j \leq d_j-1$. So the generating function for $x_j$ is
\begin{equation}
\mathcal{G}(x_j)=1+x+x^2+\cdots+x^{d_j-1}=\frac{1-x^{d_j}}{1-x}\,.
\end{equation}
Since all variables $\{x_j\}_{j=1}^n$ are independent, the generating function for their sum is
\begin{equation}
\mathcal{G}(x_1,\ldots,x_n)=\frac{\prod_{j=1}^{n}(1-x^{d_j})}{(1-x)^n}\,.
\end{equation}
To determine the coefficient of $x^k$ in the expansion of the generating function we have
\begin{align}\label{a8}
\prod_{j=1}^{n}1-x^{d_j}&=\sum_{\textsc{s}\subseteq\{1,\ldots,n\}}(-1)^{|\textsc{s}|}x^{(\sum_{j\in\textsc{s}}d_j)}\,.
\end{align}
By multiplying the two expansions in Eqs. \eqref{a4} and \eqref{a8}, and extracting the coefficient of $x^k$, we obtain the condition $m+\sum_{j\in\textsc{s}}d_j=k$, or equivalently, $m=k-\sum_{j\in\textsc{s}}d_j$ (noting that $k-\sum_{j\in\textsc{s}}d_j\geq 0$). This leads to
\begin{equation}
[x^k]\mathcal{G}(x_1,\ldots,x_n)=\sum_{\textsc{s}\subseteq\{1,\ldots,n\}}(-1)^{|\textsc{s}|}\binom{k-(\sum_{j\in\textsc{s}}d_j)+n-1}{n-1}\,.
\end{equation}
\qed

Lemma \ref{lemmaA2} is a more general result that includes Lemma \ref{lemmaA1} as a special case when $d_j=d$, for all $1\leq j\leq n$.

Notably, this quantity corresponds to the number of ways to distribute $k$ balls into $n$ bins, where the $j$-th bin ($1\leq j \leq n$) can hold at most $d_j-1$ balls. Consequently, an alternative proof can be derived using the inclusion–exclusion principle.

%%%%%%%%%%%%%%%%%%%%%%%%%%%%%%%%%%%%%%%%%%%%%%%%%

\section{Alternative algorithm}\label{AppxB}

Here, we propose an alternative algorithm that facilitates the construction of the CESs established in Theorems \ref{theorem-nqubit}-\ref{theorem-nqudit-veronese} and \ref{theorem-nqudit-SegreVeronese}-\ref{theorem-npartite}. The basis of the GES in these theorems is given as follows:
\begin{itemize}[leftmargin=2.5cm]
    \item[Theorem \ref{theorem-nqubit}:] GME Dicke states $\{|\mathrm{D}_n^k\rangle\}_{k=1}^{n-1}$,
    \item[Theorem \ref{theorem-nqudit-veronese}:] GME generalized Dicke states $\{|\mathrm{D}_n^{\bk}\rangle\mid\bk\neq\pi\{(n,0,\ldots,0)\}\}$,
    \item[Theorem \ref{theorem-nqudit-SegreVeronese}:] GME $\mathrm{G}$ states $\{|\mathrm{G}_n^k\rangle\}_{k=1}^{n(d-1)-1}$,
    \item[Theorem \ref{theorem-nqudit-VSV}:] GME $\mathscr{G}$ states $\{|\mathscr{G}_{K,n}^{s,\bs}\rangle\mid K=n:1\leq s\leq n(\mathfrak{k}+1)-1,~K=0:\bs\neq\tilde{\pi}(n,0,\ldots,0)\}$,
    \item[Theorem \ref{theorem-npartite}:] GME $\textswab{G}$ states $\{|\textswab{G}_n^k\rangle\}_{k=1}^{\textswab{S}-1}$.
\end{itemize}
These bases can be defined explicitly using Eqs. \eqref{Dicke-State}, \eqref{gen-Dicke}, \eqref{G-State}, \eqref{G-State-VSV}, and \eqref{G-State-gen}, with their corresponding normalization factors given by Eqs. \eqref{binomial}, \eqref{multinomial}, \eqref{S(n,k,d-1)}, \eqref{normalization-G-VSV}, and \eqref{S(n,k,dj-1)}.

Utilizing these normalization factors, we construct a pairwise orthonormal basis for the orthocomplement subspace CES. Let $\mathrm{N}$ denote the normalization factor of the basis elements of the GES in Theorems \ref{theorem-nqubit}-\ref{theorem-nqudit-veronese} and \ref{theorem-nqudit-SegreVeronese}-\ref{theorem-npartite}. By applying a Discrete Fourier Transform (DFT) \cite{Horn-Johnson}, we redefine the basis elements with new coefficients corresponding to the second-to-last row of the DFT matrix, given by
\begin{equation}
\mathrm{F}=\frac{1}{\sqrt{\mathrm{N}}}\Big(\omega^{(i-1)(j-1)}\Big)_{1\leq i,j\leq\mathrm{N}}.
\end{equation}
where $\omega=\exp(\frac{2\pi\bf{i}}{\mathrm{N}})$ is a primitive $\mathrm{N}$-th root of unity. Explicitly, the DFT matrix takes the form
\begin{equation}
\mathrm{F}=\frac{1}{\sqrt{\mathrm{N}}}\begin{pmatrix}
1 & 1 & 1 & \cdots & 1 \\
1 & \omega & \omega^{2} & \cdots & \omega^{(\mathrm{N}-1)} \\
1 & \omega^{2} & \omega^{4} & \cdots & \omega^{2(\mathrm{N}-1)} \\
\vdots & \vdots & \vdots & \ddots & \vdots \\
1 & \omega^{(\mathrm{N}-1)} & \omega^{2(\mathrm{N}-1)} & \cdots & \omega^{(\mathrm{N}-1)(\mathrm{N}-1)}
\end{pmatrix}.
\end{equation}
This transformation provides a systematic method for obtaining an orthonormal basis for the CES subspace, ensuring the necessary structure for our proofs.

\begin{manualexample}{B1}\label{exmp-nqubit-Appx}
Using the algorithm presented in this appendix, we obtain the following decomposition of the four-qubit Hilbert space in Example \ref{exmp-nqubit}:
\begin{align*}
\mathcal{H}_{2^4}&=\Span\{|0000\rangle,|1111\rangle\}\oplus\Span\{|\mathrm{D}_4^1\rangle,|\mathrm{D}_4^2\rangle,|\mathrm{D}_4^3\rangle\}\\
&~~ \oplus\Span\{|\phi_1^{1}\rangle=\tfrac{1}{2}(|0001\rangle+\omega_{1}|0010\rangle+\omega_{1}^{2}|0100\rangle+\omega_{1}^{3}|1000\rangle), \\
&~~\qquad\qquad |\phi_1^{2}\rangle=\tfrac{1}{2}(|0001\rangle+\omega_{1}^{2}|0010\rangle+\omega_{1}^{4}|0100\rangle+\omega_{1}^{6}|1000\rangle), \\
&~~\qquad\qquad |\phi_1^{3}\rangle=\tfrac{1}{2}(|0001\rangle+\omega_{1}^{3}|0010\rangle+\omega_{1}^{6}|0100\rangle+\omega_{1}^{9}|1000\rangle), \\
&~~\qquad\qquad |\phi_2^{1}\rangle=\tfrac{1}{\sqrt{6}}(|0011\rangle+\omega_{2}|0101\rangle+\omega_{2}^{2}|0110\rangle+\omega_{2}^{3}|1001\rangle+\omega_{2}^{4}|1010\rangle+\omega_{2}^{5}|1100\rangle), \\
&~~\qquad\qquad |\phi_2^{2}\rangle=\tfrac{1}{\sqrt{6}}(|0011\rangle+\omega_{2}^{2}|0101\rangle+\omega_{2}^{4}|0110\rangle+\omega_{2}^{6}|1001\rangle+\omega_{2}^{8}|1010\rangle+\omega_{2}^{10}|1100\rangle), \\
&~~\qquad\qquad |\phi_2^{3}\rangle=\tfrac{1}{\sqrt{6}}(|0011\rangle+\omega_{2}^{3}|0101\rangle+\omega_{2}^{6}|0110\rangle+\omega_{2}^{9}|1001\rangle+\omega_{2}^{12}|1010\rangle+\omega_{2}^{15}|1100\rangle), \\
&~~\qquad\qquad |\phi_2^{4}\rangle=\tfrac{1}{\sqrt{6}}(|0011\rangle+\omega_{2}^{4}|0101\rangle+\omega_{2}^{8}|0110\rangle+\omega_{2}^{12}|1001\rangle+\omega_{2}^{16}|1010\rangle+\omega_{2}^{20}|1100\rangle), \\
&~~\qquad\qquad |\phi_2^{5}\rangle=\tfrac{1}{\sqrt{6}}(|0011\rangle+\omega_{2}^{5}|0101\rangle+\omega_{2}^{10}|0110\rangle+\omega_{2}^{15}|1001\rangle+\omega_{2}^{20}|1010\rangle+\omega_{2}^{25}|1100\rangle), \\
&~~\qquad\qquad |\phi_3^{1}\rangle=\tfrac{1}{2}(|0111\rangle+\omega_{3}|1011\rangle+\omega_{3}^2|1101\rangle+\omega_{3}^3|1110\rangle), \\
&~~\qquad\qquad |\phi_3^{2}\rangle=\tfrac{1}{2}(|0111\rangle+\omega_{3}^{2}|1011\rangle+\omega_{3}^{4}|1101\rangle+\omega_{3}^{6}|1110\rangle), \\
&~~\qquad\qquad |\phi_3^{3}\rangle=\tfrac{1}{2}(|0111\rangle+\omega_{3}^{3}|1011\rangle+\omega_{3}^{6}|1101\rangle+\omega_{3}^{9}|1110\rangle),
\} \\
&=\Span\{|0000\rangle,|1111\rangle\}\oplus\mathrm{GES}_{3}\oplus\mathrm{CES}_{11}\,,
\end{align*}
where $\omega_1=\omega_3=\exp(\frac{\pi\bf{i}}{2})$, and $\omega_2=\exp(\frac{\pi\bf{i}}{3})$.
\end{manualexample}

Inspired by Theorem \ref{theorem-GES-GES-CES}, this algorithm enables the extraction of GESs from the CES in Theorems \ref{theorem-nqubit}-\ref{theorem-nqudit-veronese} and \ref{theorem-nqudit-SegreVeronese}-\ref{theorem-npartite}.

\begin{manualexample}{B2}\label{exmp-nqubit-Appx-2}
In Example \ref{exmp-nqubit-Appx}, one can extract three GESs of dimension three from the eleven-dimensional CES as follows:
\begin{align*}
\mathcal{H}_{2^4}=&\Span\{|0000\rangle,|1111\rangle\}\oplus\Span\{|\mathrm{D}_4^1\rangle,|\mathrm{D}_4^2\rangle,|\mathrm{D}_4^3\rangle\}\oplus\Span\{|\phi_1^1\rangle,|\phi_2^1\rangle,|\phi_3^1\rangle\}\oplus\Span\{|\phi_1^2\rangle,|\phi_2^2\rangle,|\phi_3^2\rangle\}\\
& \oplus\Span\{|\phi_1^3\rangle,|\phi_2^3\rangle,|\phi_3^3\rangle\}\oplus\Span\{|\phi_2^4\rangle,|\phi_2^5\rangle\} \\
=& \Span\{|0000\rangle,|1111\rangle\}\oplus\mathrm{GES}_{3}\oplus\mathrm{GES}_{3}\oplus\mathrm{GES}_{3}\oplus\mathrm{GES}_{3}\oplus\mathrm{CES}_{2}\,.
\end{align*}
\end{manualexample}

%%%%%%%%%%%%%%%%%%%%%%%%%%%%%%%%%%%%%%%%%%%%%%%%%

\section{Decomposition of three-qubit Hilbert space into GESs}\label{AppxC}

In the light of Theorem \ref{theorem-nqubit} and employing the algorithm introduced in Appendix \ref{AppxB}, we present a structured decomposition of a three-qubit Hilbert space $\mathcal{H}_{2^3}=\mathbbm{C}^2\otimes\mathbbm{C}^2\otimes\mathbbm{C}^2$ as follows:
\begin{align}\nonumber
\mathcal{H}_{2^3}=&\Span\{|000\rangle,|111\rangle\}\oplus\Span\{|\mathrm{D}_3^1\rangle,|\mathrm{D}_3^2\rangle\}\\ \nonumber
& \oplus\Span\{|\chi_1^{1}\rangle=\frac{1}{\sqrt{3}}(|001\rangle+\omega|010\rangle+\omega^2|100\rangle), |\chi_1^{2}\rangle=\frac{1}{\sqrt{3}}(|001\rangle+\omega^2|010\rangle+\omega|100\rangle),\\ \nonumber
&\qquad\qquad |\chi_2^{1}\rangle=\frac{1}{\sqrt{3}}(|011\rangle+\omega|101\rangle+\omega^2|110\rangle), |\chi_2^{2}\rangle=\frac{1}{\sqrt{3}}(|011\rangle+\omega^2|101\rangle+\omega|110\rangle)\} \\ \label{3qubitH}
=&\Span\{|000\rangle,|111\rangle\}\oplus\mathrm{GES}_{2}\oplus\mathrm{CES}_{4}\,,
\end{align}
where $\omega$ is a nonreal cube root of unity. Furthermore, from the four-dimensional CES, two mutually orthogonal two-dimensional GESs can be explicitly extracted, leading to the refined decomposition:
\begin{align}\nonumber
\mathcal{H}_{2^3}&=\Span\{|000\rangle,|111\rangle\}\oplus\Span\{|\mathrm{D}_3^1\rangle,|\mathrm{D}_3^2\rangle\}\oplus\Span\{|\chi_1^1\rangle,|\chi_2^2\rangle\}\oplus\Span\{|\chi_1^2\rangle,|\chi_2^1\rangle\} \\ \label{3qubitH2}
&=\Span\{|000\rangle,|111\rangle\}\oplus\mathrm{GES}_{2}\oplus\mathrm{GES}_{2}\oplus\mathrm{GES}_{2}\,.
\end{align}

The first term in the r.h.s. of Eq. \eqref{3qubitH2} can be written as
\begin{equation}
\Span\{|000\rangle,|111\rangle\}\equiv\Span\{|{\rm{GHZ}}_{+}\rangle,|{\rm{GHZ}}_{-}\rangle\}\,,
\end{equation}
with the GHZ states defined as $|{\rm{GHZ}}_{\pm}\rangle=\frac{1}{\sqrt{2}}|000\rangle\pm|111\rangle$. Since these GHZ states are GME and share no common canonical basis with the states $\{|\chi_i^j\rangle\}_{i,j=1,2}$, the decomposition in Eq.~\eqref{3qubitH2} can be rewritten as follows:
\begin{align}\nonumber
\mathcal{H}_{2^3}&=\Span\{|\mathrm{D}_3^1\rangle,|\mathrm{D}_3^2\rangle\}\oplus\Span\{|{\rm{GHZ}}_{+}\rangle,|\chi_1^1\rangle,|\chi_2^2\rangle\}\oplus\Span\{|{\rm{GHZ}}_{-}\rangle,|\chi_1^2\rangle,|\chi_2^1\rangle\} \\
&=\mathrm{GES}_{2}\oplus\mathrm{GES}_{3}\oplus\mathrm{GES}_{3}\,.
\end{align}

%%%%%%%%%%%%%%%%%%%%%%%%%%%%%%%%%%%%%%%%%%%%%%%%%
\newpage
%%%%%%%%%%%%%%%%%%%%%%%%%%%%%%%%%%%%%%%%%%%%%%%%%

%%%%%%%%%%%%%%%%%%%%%%%%%%%%%%%%%%%%%%%%%%%%%%%%%
%%%%%%%%%%%%%%%%%%%%%%%%%%%%%%%%%%%%%%%%%%%%%%%%%

\end{document}